\def\USEACHEMSO{0} %

\if\USEACHEMSO1
\documentclass[journal=jpclcd,manuscript=letter,%
email=true,%
]{achemso}
\else
\documentclass[aip,notitlepage,%
jcp,
tightenlines,
twocolumn,
reprint,%
floatfix]{revtex4-2} 
\fi

\usepackage[utf8]{inputenc}
\usepackage[T1]{fontenc} 
\usepackage{stix}
\usepackage[version=3]{mhchem}
\usepackage{hyperref}
\usepackage{units}

\usepackage{booktabs}
\usepackage{dcolumn}%

\usepackage{xspace}
\usepackage{verbatim}
\let\oldtheequation\theequation
\makeatletter
\def\tagform@#1{\maketag@@@{\ignorespaces#1\unskip\@@italiccorr}}
\renewcommand{\theequation}{(\oldtheequation)}
\makeatother 

\if\USEACHEMSO1
\fi

\usepackage{listings}

\usepackage{xcolor}
\usepackage{graphicx}

\newcommand{\matr}[1]{\ensuremath{\mathbf{#1}}}
\newcommand{\braket}[2]{\ensuremath{ \langle #1 | \, #2  \rangle }}

\newcommand{\ketbra}[2]{\ensuremath{  | {#1} \rangle \langle {#2} |}}

\newcommand{\ket}[1]{\ensuremath{  | {#1} \rangle}}

\newcommand{\matrixe}[3]{\ensuremath{ \langle{#1} | {#2} | {#3} \rangle }}

\newcommand{\icm}{cm^{-1}}

\newcommand{\lit}[1]{Ref.~\mbox{[\!\!\citenum{#1}]}\xspace}
\newcommand{\lits}[1]{Refs.~\mbox{[\!\!\citenum{#1}]}\xspace}

\definecolor{CBdblue}{RGB}{5,113,176}

\if\USEACHEMSO1
\begin{tocentry}
\includegraphics{cover/cover}%
\end{tocentry}
\fi

\if\USEACHEMSO0
\begin{document}
\fi

\title{Benchmarking vibrational spectra: 5000 accurate eigenstates of acetonitrile using tree tensor network states}
\author{Henrik R.~Larsson}
\affiliation{Department of Chemistry and Biochemistry, University of California, Merced, CA 95343, USA}

\if\USEACHEMSO1
\begin{document}
\fi

\begin{abstract}
Accurate vibrational spectra are essential for understanding how molecules behave, 
yet their computation remains challenging and benchmark data to reliably compare different methods are sparse. 
Here, we present high-accuracy eigenstate computations for the six-atom, 12-dimensional acetonitrile molecule, a prototypical, strongly coupled, anharmonic system. Using a density matrix renormalization group (DMRG) algorithm with a tree-tensor-network-state (TTNS) ansatz,
a refinement using TTNSs as basis set,
and reliable procedures to estimate energy errors, we compute up to 5,000 vibrational states with error estimates below $\unit[0.0007]{\icm}$. %
Our analysis reveals that previous works underestimated the energy  error by up to two orders of magnitude.
Our data serve as a benchmark for future vibrational spectroscopy methods and our new method %
offers a path toward similarly precise computations of large, complex molecular systems.\\
\end{abstract}

\maketitle

Vibrational spectra reveal important insights into chemical bonding, the complex interactions of atoms in molecules, and molecules in environments.\cite{Quantum2006huang,Vibrational2008wang,CH52015wodraszka,Encoding2017devine,Stateresolved2022larsson,Coupling2022schroder,Smolyak2022chen,Chromium2022larsson,Quantum2023simko}
The accurate simulation of vibrational spectra, however, is nontrivial and new methods to compute them are being developed almost on a weekly basis.\cite{Variational2008bowman,Perspective2017carrington,Computational2021barone,Exact2023matyus,Tensor2024larsson}
It would be very fruitful to assess these different methods on a common ground using benchmark problems,  
which, among others, 
have been very successful in understanding the pros and cons of electronic structure methods.\cite{Thorough2011goerigk,Solution2017motta,Mountaineering2020loos,Ground2020eriksen,Direct2020williams,Minimal2020larsson,Chromium2022larsson}
While some benchmark data exist for some specific settings in quantum vibrational dynamics,\cite{Benchmarking2017mata,LowTemperature2018welsch,Benchmark2023a.mata}
using established reference sets
of vibrational spectra 
with a well-defined error estimate to benchmark methods %
is still at its infancy.
The reason for the lack of reference data is that to date it is still extremely difficult to reliably compute accurate vibrational spectra with hundreds or even thousands of vibrational transitions in large  coupled, anharmonic molecules. 

To gauge the performance of new methods to compute vibrational spectra, %
in several dozens of studies
the vibrational spectrum of acetonitrile  (methyl cyanide, \ce{CH3CN})  has been computed.\cite{
DFT2004carbonniere, %
Calculations2005begue,%
Nonproduct2011avila,%
Using2011avila,%
Calculating2014leclerc,%
Impact2014lutz,%
Large2015halverson,%
Using2015thomas,%
Using2016browna,%
Adaptive2016garnier,%
Using2016leclerc,%
Comparison2016leclerc,%
Calculating2016rakhuba,%
Using2016wodraszka,%
Pruned2017avila,%
Vibrational2017baiardi,%
AVCI2017odunlami,%
Systematically2017wodraszka,%
Computing2019larsson,%
Dual2019garnier,%
Vibrational2019lesko,%
Lowrank2019rakhuba,%
Collocationbased2020wodraszka,%
Vibrational2021fetherolf,%
Computing2021kallullathil,%
Hitting2021sarka,%
Computing2023kallullathil,%
Computing2023simmonsa,%
Vibrational2023trana,%
Eigenstate2024hoppe,%
Using2024rey,%
Neural2024zhang}
It is  a prototypical molecule
of atmospheric, astrochemical and industrial relevance\cite{Detection1971solomon,Importance1990lobert,Infrared2012oleary} that possesses a rich and complicated vibrational spectrum, including many Fermi resonances.\cite{Measurement1999zhao,Calculations2005begue,AVCI2017odunlami,Direct2024mcdonnell}
Computing the spectrum of this six-atom, 12-dimensional vibrationally coupled, anharmonic system 
has been shown to be very challenging. For example, one of the first extensive studies missed many excited states,\cite{Using2011avila,AVCI2017odunlami,Eigenstate2024hoppe}
and several studies %
missed the targeted accuracy of the energy levels, sometimes by two orders of magnitude, %
as we will show below. 
Next to these issues, previous works on acetonitrile used vastly different settings, as they used
a different
number of computed states (15 to 10,000, see Supplementary Material (SI), Tab.~S1),
at least \emph{six different} versions of the original potential energy surface (PES)\cite{Calculations2005begue,Using2011avila,Large2015halverson,Vibrational2017baiardi,Hitting2021sarka} 
with different accuracies for the targeted states that span three orders of magnitude ($\unit[0.01]{\icm}$ to $\unit[\sim 10]{\icm}$),
and different or even no reference energies to which to compare. %
These differences 
render an objective comparison of the many developed vibrational methods difficult to impossible.
Accurate reference data on \ce{CH3CN} is needed, but current methods have difficulties with computing \emph{many} eigenstates to \emph{high} accuracy.

Our contribution in this letter is twofold. %
In our first contribution,
we report a novel computational methodology to compute thousands of vibrational states with a dramatically high accuracy.
Importantly, our approach provides reliable error estimates. 
In our second contribution, we report, for the first time,
the computation of 1,000 to 5,000 vibrational states for three different PESs of \ce{CH3CN} with an estimated error below $\unit[0.0007]{\icm}$.
Compared to existing data for \ce{CH3CN}, our new computational method leads to a fourfold
increase of the number of computed states and, simultaneously, a boost in accuracy by two orders of magnitude.
We analyze the computed states using concepts from quantum information theory\cite{First2012westermann,Vibrational2024glaser} and compare our highly accurate energies with existing data, where we show that the accuracies of some states previously have been vastly overestimated by two orders of magnitude.

\begin{figure}
  \includegraphics[width=.4\textwidth]{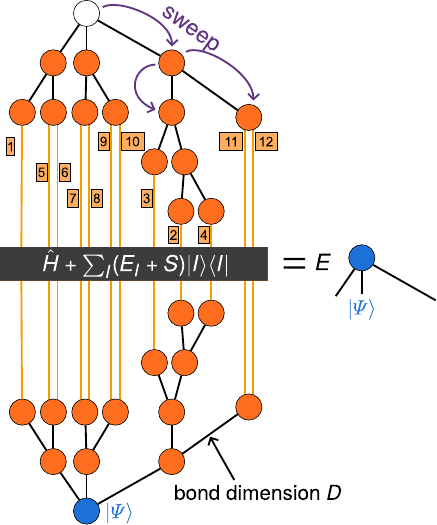}
  \caption{Tensor network diagram of the variational DMRG optimization. Nodes correspond to tensors with orange (physical) and black (virtual) vertices denoting their respective dimensions; shared vertices indicate contractions, while the free-standing vertices remain uncontracted.
  The ordering of the physical dimensions (normal modes) is shown in the orange boxes.
  The large black tensor pictorially represents the shifted Hamiltonian, which here is not shown as a full tensor network. 
  On the left, a matrix-eigenvector product arises from contracting the effective Hamiltonian (orange and black tensors) with the blue tensor  $\ket{\Psi}$. Once optimized, the new blue tensor replaces its predecessor in the TTNS (empty circle). 
  This procedure is repeated for all tensors in one ``sweep'' (first steps shown in purple).
  }
  \label{fig:ttns}
\end{figure}

Our method is based on the density matrix renormalization group (DMRG)\cite{Density1992white,Practical2014orus} applied to vibrational tree tensor network states (TTNSs),\cite{Computing2019larsson} which is the underlying ansatz of the %
multilayer multiconfigurational time-dependent Hartree method (ML-MCTDH).\cite{Studying2012meyer,Multilayer2015wang,Wavepacket2017manthe,Tensor2024larsson}
The resulting algorithm is schematically shown in  \autoref{fig:ttns}.
To compute excited states, we use the state-shifted Hamiltonian\cite{Iterative1973shavitt,Computing2019larsson}  $\hat H + \sum_I (E_I + S) \ketbra{I}{I}$, where $S$ is a number large enough to ensure separation between the shifted states $\ket{I}$ with energy $E_I$ and the next lowest-lying state.
While we have used this procedure to compute up to $2500$ states with errors below $\unit[1]{\icm}$,\cite{Stateresolved2022larsson,vibronic2024larsson}
the computed states can become slightly non-orthogonal with overlaps on the order of $|\braket{I}{J}|\sim10^{-6}$, which leads to an energy error of $\unit[0.2]{\icm}$.
To avoid this error, 
we here refine our states by diagonalizing our Hamiltonian in the basis of computed TTNSs: After we compute the necessary number of TTNS states %
using the DMRG, we compute the Hamiltonian matrices $H_{IJ}=\matrixe{I}{\hat H}{J}$ and  overlap matrices $S_{IJ}=\braket{I}{J}$, and solve the generalized eigenvalue problem, $\matr H \matr U = \matr S \matr U \matr E$, where $\matr E$ contains the final eigenvalues. 
The computation of the  matrix elements is embarrassingly parallel and the final matrix diagonalization has a negligible runtime on a single CPU. %
Thus, compared to the DMRG optimizations the refinement step takes significantly less runtime, but it dramatically reduces the error of the eigenstates.
Note that the refinement leads to eigenstates that are linear combinations of TTNSs. In practice, this does not cause issues as most coefficients are negligible. We use DMRG-like procedures\cite{Practical2014orus,Block22023zhai} to fit the linear combination of TTNSs to one final TTNS. We then use the fitted TTNSs to conveniently compute observables. %

Error estimates are crucial to reliably compute states to high accuracy.
For a given Hamiltonian, the two main numerical errors our method introduces are both due to finite bases.
The first one is due to the physical basis that discretizes the coordinates,
and the second one is due to the ``virtual'' renormalized basis
represented by the tensors in the TTNS, whose finite size is called bond dimension, $D$. The maximum value of $D$ is dubbed here $D_\text{max}$. 
MCTDH users call $D$ ``number of single-particle functions.''\cite{Tensor2024larsson}

To render the finite basis-set error negligible, we use a very large physical basis with 42 Gauß-Hermite DVR functions\cite{Discretevariable2000light,PhaseSpace2018tannor}
in each dimension.
We can use this large basis, as it is not a bottleneck in the TTNS computation.
However, to improve the sparsity in our Hamiltonian we use the DVR approximation, which results in a diagonal potential energy matrix.\cite{Efficient2016larsson,Dynamical2017larsson}
To estimate the physical basis error and the DVR error, %
we convert our final TTNSs from the DVR basis to a fully variational Harmonic oscillator basis with 10\% fewer functions than DVR points, $\{\ket{\text{HO}_h}\}_{h=1}^{38}$,
and use the difference between the DVR and the variational energies as error estimate. We change the basis by applying the projector $\sum_h \ketbra{\text{HO}_h}{\text{HO}_h}$ %
in each dimension.
For systems with very complex PESs, this procedure is too costly but can be adjusted by evaluating the energy on a DVR grid with fewer basis functions.

\begin{figure}
  \includegraphics{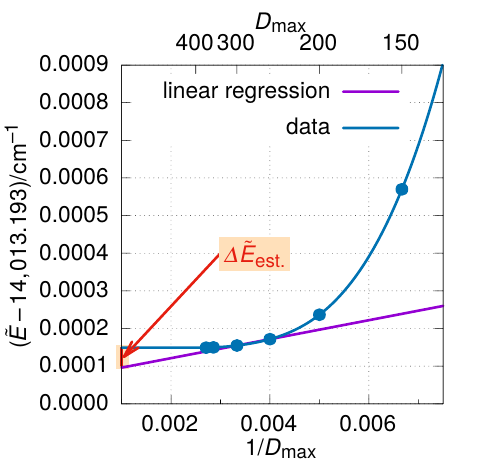}
  \caption{Example of the calculation of the DMRG error estimate. The data points depict DMRG energies at different inverse max.~bond dimensions $1/D_\text{max}$, and the blue line is based on a cubic spline interpolation.
  The purple line is based on a linear fit of two data points. The error estimate then is the difference of the best energy and the linear fit at $1/D_\text{max}=0$.
  The upper abscissa shows $D_\text{max}$, for comparison.
  Here, for better visualization the linear fit is performed for two smaller $D_\text{max}$ values,  leading to an error estimate that is larger than necessary.
  The data is based on state 1002 optimized on the AC PES.\cite{Using2011avila}}
  \label{fig:Dextrapol}
\end{figure}

Our TTNS approach adapts the bond dimensions dynamically
during the DMRG optimization by discarding all singular values of the TTNS tensors below a threshold %
using the procedure described in \lit{Computing2019larsson}.
This leads to a different $D_\text{max}$ for each computed TTNS.
We estimate the resulting finite basis error %
using two separate procedures. The first one is based on the assumption that the energy as a function of $1/D_\text{max}$ is convex.\cite{Simulation2009tagliacozzo} 
A linear extrapolation of two energies computed using states with different $D_\text{max}$ values then provides a lower bound of the energy, see \autoref{fig:Dextrapol} for an example.
Since our bond dimension adaption not only leads to a  different $D_\text{max}$ for each state but also
affects all virtual bases, including those that have a bond dimension smaller than $D_\text{max}$, 
here, after the initial DMRG optimization we first re-optimize each state by allowing each bond dimension to grow as large as $D_\text{max}$ regardless of the singular values. 
Then, we repeat the DMRG optimization using a max.~bond dimension of $0.8 D_\text{max}$ %
and take the difference between the lower energy bound from the linear regression and the lowest energy as first error estimate. 
Note that this does not take spurious orthogonality errors into account. To include them, in our second estimate we first compress each state using singular value decomposition to get bond dimensions up to $0.8 D_\text{max}$, and then solve another generalized eigenvalue problem in the basis of compressed TTNSs. The second estimate is then calculated in the same way as in the first approach but using the energies from the generalized eigenvalue problem. 
We use the larger value of these two estimates as final estimate of the finite  $D_\text{max}$ error.
Note that other error estimates exist for the DMRG.%
\cite{Error2018hubig,Tensor2024larsson} 
These  dominantly take errors due to two-body correlation and not higher-order correlations into account, %
and, 
for our application, are more resource-intensive to evaluate than our proposed procedures. 

Since we do not symmetry-adapt our basis, deviations in the energy of degenerate $E$-symmetric ($C_{3v}$ point group) states may occur. We estimate this degeneracy error by the energy difference of degenerate states.
We then evaluate the total error as the square root of the sum of the squared individual error contributions.

To be consistent with previous computations on \ce{CH3CN}, we use the simplified $J=0$ Hamiltonian for normal coordinates %
that as kinetic energy operator uses 
$- 0.5 \sum_\kappa \omega_\kappa\ \partial^2/ \partial \hat q_\kappa^2$, 
where $\hat q_\kappa$ is the position operator of mode $\kappa$ and $\omega_\kappa$ is its angular frequency.
We use the TTNS structure shown in  \autoref{fig:ttns}.
Most vibrational spectra computations of \ce{CH3CN} use a PES that is based on the quartic Taylor expansion of \lit{Calculations2005begue}, where only the largest coefficients are reported for an 8-dimensional subset of the 12-dimensional potential.
While the full 12-dimensional potential can be retrieved from symmetry relations,\cite{Cubic1961henry,Quartic1965henry} ambiguities of this procedure led to at least five versions of the original PES in \lit{Calculations2005begue}.\cite{Using2011avila,Large2015halverson,Vibrational2017baiardi,Hitting2021sarka}
Here, we will use the most commonly used version 
by Avila and Carrington (AC),\cite{Using2011avila} %
where up to 240 states have been reported,
and two additional versions by Sarka and Poririer (CSC, USC), who provided energies for the first 1000 states to which we can compare.\cite{Hitting2021sarka}
The USC PES is very similar to the AC PES whereas the CSC PES leads to less correlated states, as we will show below.
We note that our computations are far more  accurate than the quality of the used PESs. Similar to related electronic structure benchmarks,
our computed energies thus are mostly for benchmark purposes. 
We chose the PES due to its ubiquitous use in the literature, and we are not aware of a PES for acetonitrile that has a better quality.
However, our procedure is fully transferable to more realistic Hamiltonians. 
See the SI for further details on the error estimate and on the simulation parameters.

\begin{figure}
  \includegraphics{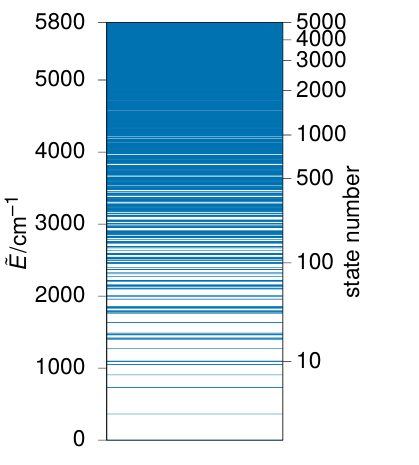}
  \caption{Computed excitation energy levels on the CSC PES. A line is plotted for each level. The left (right) ordinate displays the wavenumber (state number).
  Due to the high density of states, individual states cannot be identified once 500 states are reached.
  The levels for the USC and AC PESs followed the same trend.
  }
  \label{fig:horizontal_spec}
\end{figure}

\begin{figure}
  \includegraphics{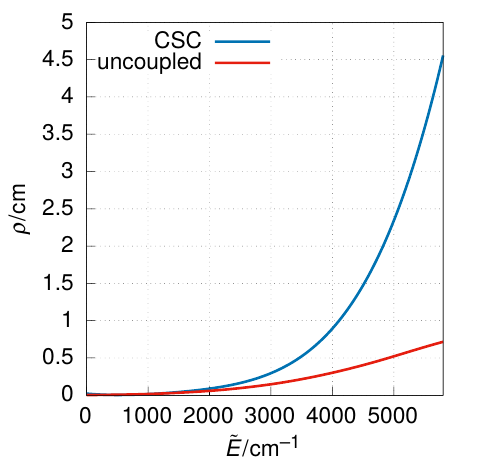}
  \caption{Density of states (DOS) for the CSC PES and an uncoupled PES where all couplings between effective modes have been removed.
  The DOS is based on a 5th-order polynomial fit for the CSC PES and a Gaussian kernel density estimate for the uncoupled PES.
  The DOS for the USC and AC PESs followed the CSC DOS.}
  \label{fig:dos}
\end{figure}

\autoref{fig:horizontal_spec} gives an overview of the computed energy levels. With 5000 computed eigenstates on the CSC PES, we reach the $\unit[5800]{\icm}$ excitation energy mark. %
As is typical for an
anharmonic, coupled vibrational systems, 
the density of states increases rapidly after the first $\sim$100 states are reached. %
For comparison, there are 100 states up to $\unit[2470]{\icm}$, %
but already approximately 2000 states up to $\unit[4850]{\icm}$.
The density of states as a function of excitation energy is shown in \autoref{fig:dos} (blue curve) and displays a steep increase of the number of states from $\sim 0.5$ states per $\unit{\icm}$ at $\unit[3500]{\icm}$ to almost $4.5$ states   per $\unit{\icm}$ at $\unit[5800]{\icm}$. In contrast, the density of states for an uncoupled model of \ce{CH3CN} (red curve in \autoref{fig:dos}) increases modestly and leads to only $0.7$ states per $\unit{\icm}$ at  $\unit[5800]{\icm}$.

For the 1000 computed states on the USC and AC PESs, we reach an excitation energy of $\unit[4174]{\icm}$.
The spectrum is very similar to that of the CSC PES, but
we found 2 (3) eigenstates localized in unphysical holes for the AC (USC) PES (the number of holes is limited by the range of the DVR basis), which we excluded from our analysis.   
In agreement with a previous analysis,\cite{Hitting2021sarka}
we did not find any holes for the CSC PES.
Notably, the holes in the AC and USC PESs did not cause numerical issues for our TTNS method. %
The small error estimates (see below) and their insensitivity to different bases indicate that the holes do not significantly distort other states.

\begin{figure*}
  \includegraphics{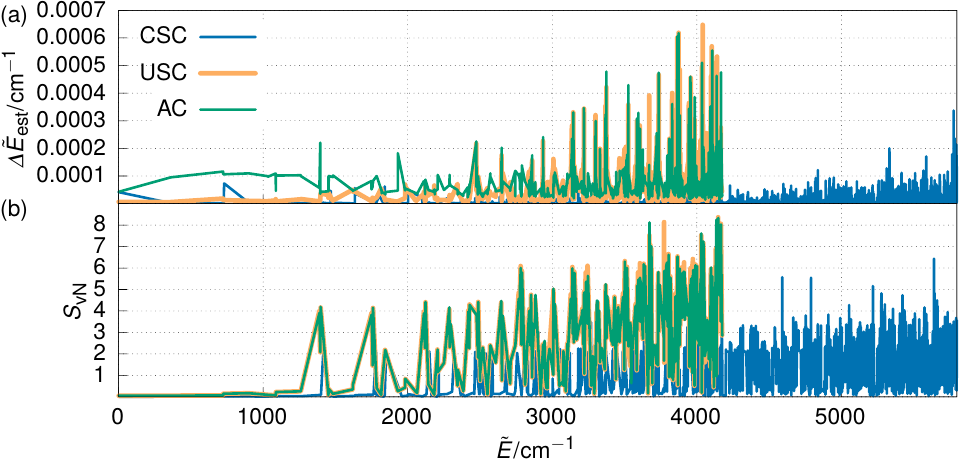}
  \caption{Error estimate (a) and entanglement/von Neumann entropy (b) for the three PESs. The data points for each individual state are connected. The data values for the USC and AC PESs are very similar.}
  \label{fig:errorEstimate_Entropy}
\end{figure*}

\autoref{fig:errorEstimate_Entropy}(a) depicts the error estimates for  all eigenstate computations on the three used PESs.
The error estimates for the AC/USC PESs are well-below  $\unit[0.0007]{\icm}$ for  all the computed 1000 states.
The individual error contributions are shown in the SI (Figs.~S1-S3). In all cases, the error due to finite $D_\text{max}$ is dominating, whereas the DVR and the degeneracy errors are negligible.
The TTNS refinement procedure dramatically reduces the degeneracy error, e.g.~for the CSC PES from $\unit[0.03]{\icm}$ to $\unit[0.00004]{\icm}$.
Remarkably, the CSC energies are much more accurate for all 5000 states and the error estimate just reaches values around $\unit[0.0002]{\icm}$ for the last few of the 5000 eigenstates.
Why is the error for the CSC PES so much smaller than that for the AC/USC PES?
We can answer this by analyzing the entanglement (von Neumann) entropy shown in \autoref{fig:errorEstimate_Entropy}(b).
Here, the entanglement entropy is defined as the sum of single-modal (or single-particle) entropies,\cite{First2012westermann,Vibrational2024glaser}
which is obtained through the reduced density operators $\hat\rho^{(\kappa)}$ in each dimension $\kappa$;
$S_{\text{vN}}=-\sum_{\kappa}  \text{tr}[ \hat\rho^{(\kappa)} \ln \hat\rho^{(\kappa)}]$. 
A definition based on a bipartition of the TTNS\cite{Matrix2022larsson} leads to similar qualitative results.
A value of $S_{\text{vN}}=0$ indicates a pure product state.
For the first 1000 states,
the entropies of the AC/USC states are in some cases larger than $8$ whereas those of the CSC PES are all below $4$. %
Since the entanglement entropy can be regarded as measure of correlation, %
the CSC PES thus leads to much less-correlated states than the AC/USC PESs. As correlated states are more difficult to describe, the error for the AC/USC states is larger.
The jaggedness of the entropies as function of energy is due to the different nature of each state.
We find that states with small entropies have simple excitation patterns that can be reasonably well approximated by product states, whereas  most of the
states with large entropies  are Fermi resonances or show other types of strong modal coupling.

The maximum bond dimension for each state %
is another measure of modal coupling and correlation (see Fig.~S4 in the SI). To reach the targeted error estimate, 
for the AC/USC PESs bond dimensions of almost $450$ are required
whereas for the CSC PES bond dimensions of up to $200$ are sufficient. We note that these bond dimensions are an order of magnitude larger than what is typically used in vibrational dynamics. This is required to reach our targeted level of accuracy. 

\begin{figure*}
  \includegraphics{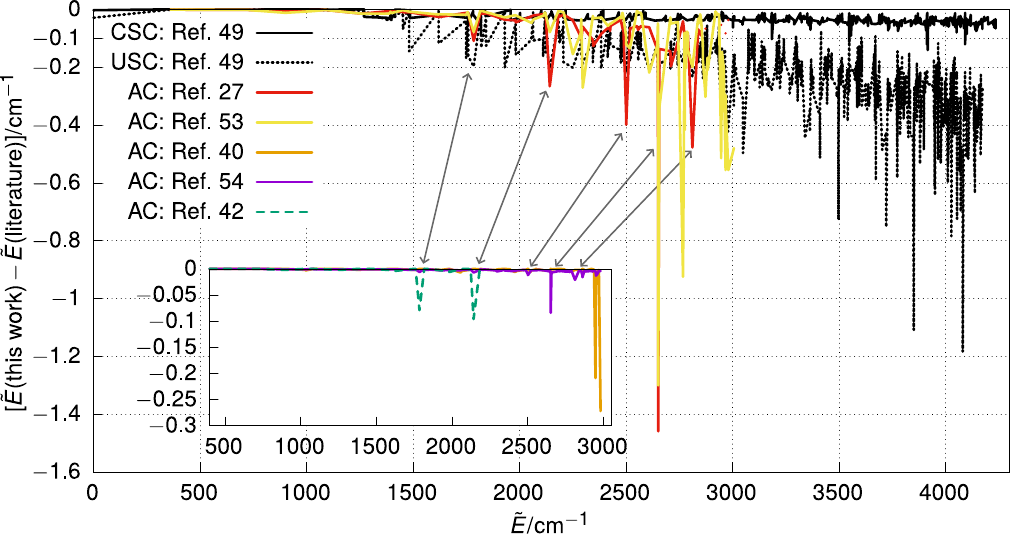}
  \caption{Difference of a subset of our computed values with selected results from the literature. A negative value means that our quasi-variational energies are more accurate.
  To improve clarity, some of the data are plotted in the inset.
  Arrows compare common outliers on the AC PES.
  }%
  \label{fig:litErr}
\end{figure*}

How do our computed energies compare with those from the literature?
Since there are at least 30 different reported computations in the literature (see Tab.~S1 in the SI), we here only compare to some selected ones. 
Further, since we here report, to the best of our knowledge, the largest number of accurately computed states for \ce{CH3CN}, we can only compare to the
smaller subset of computed states available in  the literature.
Specifically, we compare with the first 1000 computed states on the CSC and USC PESs from \lit{Hitting2021sarka}
and with the first 80 to 240 states on the AC PES from \lit{Using2011avila,AVCI2017odunlami,Computing2019larsson,Eigenstate2024hoppe,Using2024rey}, which includes results from our previous, less accurate TTNS methodology.\cite{Computing2019larsson}
\autoref{fig:litErr} shows
the difference of a subset of our energy levels with those from the literature. 
Since our DVR error is negligible, all our computed states can be regarded as variational, and  
all of our energies are below the reported literature values. This confirms the accuracy of our energies. %

The AC PES energy errors from \lit{AVCI2017odunlami} have two outliers. %
We have performed additional TTNS computations with the basis used in \lit{AVCI2017odunlami} that show that these two outliers are due to a too small basis, which is, in fact, the major error contributor for the states computed in \lit{AVCI2017odunlami} (see Fig.~S5 in the SI).
Notably, the remaining  \lits{Using2011avila,Computing2019larsson,Eigenstate2024hoppe,Using2024rey} share many outliers,
particularly, %
states at $1785$, $2142$, $2501$ and $\unit[2652]{\icm}$ (\lit{Computing2019larsson} only computed the first two)
highlighted in \autoref{fig:litErr}.
These four states all have in common that they consist of two- or three-fold excitations in mode 4. Surprisingly, these are relatively simple states with some coupling to modes 2 and 3, but they are not part of Fermi resonances and have low entanglement entropies. 
While the large errors for these states from our pioneering TTNS computation in \lit{Computing2019larsson} are due to a small DVR basis, the reason for these large errors in the remaining literature is not fully clear, but might be attributed to a strong mixing of overtones.\cite{Using2024rey}

Save for a few outliers and high-energy states that were missed in \lit{Using2011avila}, %
our energy differences to the literature values mostly agree with the convergence tests reported in the literature for the AC PES.  
This differs for the CSC and USC PES where in \lit{Hitting2021sarka} ``an overall numerical convergence accuracy of $\unit[10^{-2}]{\icm}$ or better`` %
is reported.
Strikingly, our results show that the energy error of  \lit{Hitting2021sarka} is two orders of magnitude larger:
the largest difference of our values to theirs for the CSC PES is $\unit[0.14]{\icm}$ and for the USC PES even $\unit[1.18]{\icm}$,
despite the careful convergence tests performed in \lit{Hitting2021sarka}.
This further demonstrates how difficult it is to reliably and accurately compute a large manifold of vibrational eigenstates for coupled, anharmonic molecules.

In conclusion, we here presented not only a method based on TTNSs, the DMRG, and a diagonalization-based refinement that enables the accurate computation of several thousands of vibrational eigenstates, but also a  reliable means to estimate their energy error.
We applied these new methods to  the six-atom \ce{CH3CN} molecule.
Compared to previous works, our results for \ce{CH3CN} increase the number of computed states by up to a factor of 5 and simultaneously increase the accuracy of these states by a factor of more than  $140$. %
Our comparison to existing data revealed that not all previously reported energies have the accuracy estimated in the respective literature. 
Our reported energies provide benchmark data for future comparisons to new vibrational methods for this prototypical, coupled, anharmonic molecule.
Notably,
using existing machinery to set up Hamiltonians for TTNSs,\cite{Automatic2012ndong,Transforming2020schroder,Stateresolved2022larsson,vibronic2024larsson}
our work is directly transferable to other challenging vibrational systems such as fluxional molecules,
and 
it paves the way for reliable ro-vibronic computations.
This work focused solely on computing the eigenstates and their errors. %
Our lab is currently working on automatically analyzing them to provide new insights into the intricate dynamics of strongly coupled vibrational modes as well as to discover new physical mechanisms.
\section*{Associated Content}
Supporting Information available: 
Computational details, additional data on the \ce{CH3CN} literature and on errors and bond dimensions, full lists of energies and observables, and used force field parameters. 
\if\USEACHEMSO1
\begin{acknowledgement}
\else
\acknowledgements
\fi
We thank Tucker Carrington and Uwe Manthe for helpful discussions. 
We thank Tucker Carrington for providing us with the AC PES. 
This work was supported by the US
National Science Foundation (NSF) via grant no.~CHE-2312005.
Additional support came from University of California, Merced, start-up funding, %
and through
computational time on the Pinnacles and Merced clusters at University of California, Merced (supported by NSF OAC-2019144 and ACI-1429783).
\if\USEACHEMSO1
\end{acknowledgement}
\fi


\begin{thebibliography}{81}%
\makeatletter
\providecommand \@ifxundefined [1]{%
 \@ifx{#1\undefined}
}%
\providecommand \@ifnum [1]{%
 \ifnum #1\expandafter \@firstoftwo
 \else \expandafter \@secondoftwo
 \fi
}%
\providecommand \@ifx [1]{%
 \ifx #1\expandafter \@firstoftwo
 \else \expandafter \@secondoftwo
 \fi
}%
\providecommand \natexlab [1]{#1}%
\providecommand \enquote  [1]{``#1''}%
\providecommand \bibnamefont  [1]{#1}%
\providecommand \bibfnamefont [1]{#1}%
\providecommand \citenamefont [1]{#1}%
\providecommand \href@noop [0]{\@secondoftwo}%
\providecommand \href [0]{\begingroup \@sanitize@url \@href}%
\providecommand \@href[1]{\@@startlink{#1}\@@href}%
\providecommand \@@href[1]{\endgroup#1\@@endlink}%
\providecommand \@sanitize@url [0]{\catcode `\\12\catcode `\$12\catcode
  `\&12\catcode `\#12\catcode `\^12\catcode `\_12\catcode `\%12\relax}%
\providecommand \@@startlink[1]{}%
\providecommand \@@endlink[0]{}%
\providecommand \url  [0]{\begingroup\@sanitize@url \@url }%
\providecommand \@url [1]{\endgroup\@href {#1}{\urlprefix }}%
\providecommand \urlprefix  [0]{URL }%
\providecommand \Eprint [0]{\href }%
\providecommand \doibase [0]{https://doi.org/}%
\providecommand \selectlanguage [0]{\@gobble}%
\providecommand \bibinfo  [0]{\@secondoftwo}%
\providecommand \bibfield  [0]{\@secondoftwo}%
\providecommand \translation [1]{[#1]}%
\providecommand \BibitemOpen [0]{}%
\providecommand \bibitemStop [0]{}%
\providecommand \bibitemNoStop [0]{.\EOS\space}%
\providecommand \EOS [0]{\spacefactor3000\relax}%
\providecommand \BibitemShut  [1]{\csname bibitem#1\endcsname}%
\let\auto@bib@innerbib\@empty
%
\bibitem [{\citenamefont {Huang}\ \emph {et~al.}(2006)\citenamefont {Huang},
  \citenamefont {McCoy}, \citenamefont {Bowman}, \citenamefont {Johnson},
  \citenamefont {Savage}, \citenamefont {Dong},\ and\ \citenamefont
  {Nesbitt}}]{Quantum2006huang}%
  \BibitemOpen
  \bibfield  {author} {\bibinfo {author} {\bibfnamefont {X.}~\bibnamefont
  {Huang}}, \bibinfo {author} {\bibfnamefont {A.~B.}\ \bibnamefont {McCoy}},
  \bibinfo {author} {\bibfnamefont {J.~M.}\ \bibnamefont {Bowman}}, \bibinfo
  {author} {\bibfnamefont {L.~M.}\ \bibnamefont {Johnson}}, \bibinfo {author}
  {\bibfnamefont {C.}~\bibnamefont {Savage}}, \bibinfo {author} {\bibfnamefont
  {F.}~\bibnamefont {Dong}},\ and\ \bibinfo {author} {\bibfnamefont {D.~J.}\
  \bibnamefont {Nesbitt}},\ }\bibfield  {title} {\enquote {\bibinfo {title}
  {Quantum {{Deconstruction}} of the {{Infrared Spectrum}} of
  {{CH}}{\textsubscript{5}}{\textsuperscript{+}}},}\ }\href
  {https://doi.org/10.1126/science.1121166} {\bibfield  {journal} {\bibinfo
  {journal} {Science}\ }\textbf {\bibinfo {volume} {311}},\ \bibinfo {pages}
  {60--63} (\bibinfo {year} {2006})}\BibitemShut {NoStop}%
\bibitem [{\citenamefont {Wang}\ and\ \citenamefont
  {Carrington}(2008)}]{Vibrational2008wang}%
  \BibitemOpen
  \bibfield  {author} {\bibinfo {author} {\bibfnamefont {X.-G.}\ \bibnamefont
  {Wang}}\ and\ \bibinfo {author} {\bibfnamefont {T.}~\bibnamefont
  {Carrington}},\ }\bibfield  {title} {\enquote {\bibinfo {title} {Vibrational
  energy levels of {{CH}}{\textsubscript{5}}{\textsuperscript{+}}},}\ }\href
  {https://doi.org/10.1063/1.3027825} {\bibfield  {journal} {\bibinfo
  {journal} {J. Chem. Phys.}\ }\textbf {\bibinfo {volume} {129}},\ \bibinfo
  {pages} {234102} (\bibinfo {year} {2008})}\BibitemShut {NoStop}%
\bibitem [{\citenamefont {Wodraszka}\ and\ \citenamefont
  {Manthe}(2015)}]{CH52015wodraszka}%
  \BibitemOpen
  \bibfield  {author} {\bibinfo {author} {\bibfnamefont {R.}~\bibnamefont
  {Wodraszka}}\ and\ \bibinfo {author} {\bibfnamefont {U.}~\bibnamefont
  {Manthe}},\ }\bibfield  {title} {\enquote {\bibinfo {title}
  {{{CH}}{\textsubscript{5}}{\textsuperscript{+}}: {{Symmetry}} and the
  {{Entangled Rovibrational Quantum States}} of a {{Fluxional Molecule}}},}\
  }\href {https://doi.org/10.1021/acs.jpclett.5b01869} {\bibfield  {journal}
  {\bibinfo  {journal} {J. Phys. Chem. Lett.}\ }\textbf {\bibinfo {volume}
  {6}},\ \bibinfo {pages} {4229--4232} (\bibinfo {year} {2015})}\BibitemShut
  {NoStop}%
\bibitem [{\citenamefont {DeVine}\ \emph {et~al.}(2017)\citenamefont {DeVine},
  \citenamefont {Weichman}, \citenamefont {Laws}, \citenamefont {Chang},
  \citenamefont {Babin}, \citenamefont {Balerdi}, \citenamefont {Xie},
  \citenamefont {Malbon}, \citenamefont {Lineberger}, \citenamefont {Yarkony},
  \citenamefont {Field}, \citenamefont {Gibson}, \citenamefont {Ma},
  \citenamefont {Guo},\ and\ \citenamefont {Neumark}}]{Encoding2017devine}%
  \BibitemOpen
  \bibfield  {author} {\bibinfo {author} {\bibfnamefont {J.~A.}\ \bibnamefont
  {DeVine}}, \bibinfo {author} {\bibfnamefont {M.~L.}\ \bibnamefont
  {Weichman}}, \bibinfo {author} {\bibfnamefont {B.}~\bibnamefont {Laws}},
  \bibinfo {author} {\bibfnamefont {J.}~\bibnamefont {Chang}}, \bibinfo
  {author} {\bibfnamefont {M.~C.}\ \bibnamefont {Babin}}, \bibinfo {author}
  {\bibfnamefont {G.}~\bibnamefont {Balerdi}}, \bibinfo {author} {\bibfnamefont
  {C.}~\bibnamefont {Xie}}, \bibinfo {author} {\bibfnamefont {C.~L.}\
  \bibnamefont {Malbon}}, \bibinfo {author} {\bibfnamefont {W.~C.}\
  \bibnamefont {Lineberger}}, \bibinfo {author} {\bibfnamefont {D.~R.}\
  \bibnamefont {Yarkony}}, \bibinfo {author} {\bibfnamefont {R.~W.}\
  \bibnamefont {Field}}, \bibinfo {author} {\bibfnamefont {S.~T.}\ \bibnamefont
  {Gibson}}, \bibinfo {author} {\bibfnamefont {J.}~\bibnamefont {Ma}}, \bibinfo
  {author} {\bibfnamefont {H.}~\bibnamefont {Guo}},\ and\ \bibinfo {author}
  {\bibfnamefont {D.~M.}\ \bibnamefont {Neumark}},\ }\bibfield  {title}
  {\enquote {\bibinfo {title} {Encoding of vinylidene isomerization in its
  anion photoelectron spectrum},}\ }\href
  {https://doi.org/10.1126/science.aao1905} {\bibfield  {journal} {\bibinfo
  {journal} {Science}\ }\textbf {\bibinfo {volume} {358}},\ \bibinfo {pages}
  {336--339} (\bibinfo {year} {2017})}\BibitemShut {NoStop}%
\bibitem [{\citenamefont {Larsson}\ \emph
  {et~al.}(2022{\natexlab{a}})\citenamefont {Larsson}, \citenamefont
  {Schr{\"o}der}, \citenamefont {Beckmann}, \citenamefont {Brieuc},
  \citenamefont {Schran}, \citenamefont {Marx},\ and\ \citenamefont
  {Vendrell}}]{Stateresolved2022larsson}%
  \BibitemOpen
  \bibfield  {author} {\bibinfo {author} {\bibfnamefont {H.~R.}\ \bibnamefont
  {Larsson}}, \bibinfo {author} {\bibfnamefont {M.}~\bibnamefont
  {Schr{\"o}der}}, \bibinfo {author} {\bibfnamefont {R.}~\bibnamefont
  {Beckmann}}, \bibinfo {author} {\bibfnamefont {F.}~\bibnamefont {Brieuc}},
  \bibinfo {author} {\bibfnamefont {C.}~\bibnamefont {Schran}}, \bibinfo
  {author} {\bibfnamefont {D.}~\bibnamefont {Marx}},\ and\ \bibinfo {author}
  {\bibfnamefont {O.}~\bibnamefont {Vendrell}},\ }\bibfield  {title} {\enquote
  {\bibinfo {title} {State-resolved infrared spectrum of the protonated water
  dimer: Revisiting the characteristic proton transfer doublet peak},}\ }\href
  {https://doi.org/10.1039/D2SC03189B} {\bibfield  {journal} {\bibinfo
  {journal} {Chem. Sci.}\ }\textbf {\bibinfo {volume} {13}},\ \bibinfo {pages}
  {11119--11125} (\bibinfo {year} {2022}{\natexlab{a}})}\BibitemShut {NoStop}%
\bibitem [{\citenamefont {Schr{\"o}der}\ \emph {et~al.}(2022)\citenamefont
  {Schr{\"o}der}, \citenamefont {Gatti}, \citenamefont {Lauvergnat},
  \citenamefont {Meyer},\ and\ \citenamefont
  {Vendrell}}]{Coupling2022schroder}%
  \BibitemOpen
  \bibfield  {author} {\bibinfo {author} {\bibfnamefont {M.}~\bibnamefont
  {Schr{\"o}der}}, \bibinfo {author} {\bibfnamefont {F.}~\bibnamefont {Gatti}},
  \bibinfo {author} {\bibfnamefont {D.}~\bibnamefont {Lauvergnat}}, \bibinfo
  {author} {\bibfnamefont {H.-D.}\ \bibnamefont {Meyer}},\ and\ \bibinfo
  {author} {\bibfnamefont {O.}~\bibnamefont {Vendrell}},\ }\bibfield  {title}
  {\enquote {\bibinfo {title} {The coupling of the hydrated proton to its first
  solvation shell},}\ }\href {https://doi.org/10.1038/s41467-022-33650-w}
  {\bibfield  {journal} {\bibinfo  {journal} {Nat. Commun.}\ }\textbf {\bibinfo
  {volume} {13}},\ \bibinfo {pages} {6170} (\bibinfo {year}
  {2022})}\BibitemShut {NoStop}%
\bibitem [{\citenamefont {Chen}\ \emph {et~al.}(2022)\citenamefont {Chen},
  \citenamefont {Benoit}, \citenamefont {Scribano}, \citenamefont {Nauts},\
  and\ \citenamefont {Lauvergnat}}]{Smolyak2022chen}%
  \BibitemOpen
  \bibfield  {author} {\bibinfo {author} {\bibfnamefont {A.}~\bibnamefont
  {Chen}}, \bibinfo {author} {\bibfnamefont {D.~M.}\ \bibnamefont {Benoit}},
  \bibinfo {author} {\bibfnamefont {Y.}~\bibnamefont {Scribano}}, \bibinfo
  {author} {\bibfnamefont {A.}~\bibnamefont {Nauts}},\ and\ \bibinfo {author}
  {\bibfnamefont {D.}~\bibnamefont {Lauvergnat}},\ }\bibfield  {title}
  {\enquote {\bibinfo {title} {Smolyak {{Algorithm Adapted}} to a
  {{System}}--{{Bath Separation}}: {{Application}} to an {{Encapsulated
  Molecule}} with {{Large-Amplitude Motions}}},}\ }\href
  {https://doi.org/10.1021/acs.jctc.2c00108} {\bibfield  {journal} {\bibinfo
  {journal} {J. Chem. Theory Comput.}\ }\textbf {\bibinfo {volume} {18}},\
  \bibinfo {pages} {4366--4372} (\bibinfo {year} {2022})}\BibitemShut {NoStop}%
\bibitem [{\citenamefont {Larsson}\ \emph
  {et~al.}(2022{\natexlab{b}})\citenamefont {Larsson}, \citenamefont {Zhai},
  \citenamefont {Umrigar},\ and\ \citenamefont {Chan}}]{Chromium2022larsson}%
  \BibitemOpen
  \bibfield  {author} {\bibinfo {author} {\bibfnamefont {H.~R.}\ \bibnamefont
  {Larsson}}, \bibinfo {author} {\bibfnamefont {H.}~\bibnamefont {Zhai}},
  \bibinfo {author} {\bibfnamefont {C.~J.}\ \bibnamefont {Umrigar}},\ and\
  \bibinfo {author} {\bibfnamefont {G.~K.-L.}\ \bibnamefont {Chan}},\
  }\bibfield  {title} {\enquote {\bibinfo {title} {The {{Chromium Dimer}}:
  {{Closing}} a {{Chapter}} of {{Quantum Chemistry}}},}\ }\href
  {https://doi.org/10.1021/jacs.2c06357} {\bibfield  {journal} {\bibinfo
  {journal} {J. Am. Chem. Soc.}\ }\textbf {\bibinfo {volume} {144}},\ \bibinfo
  {pages} {15932--15937} (\bibinfo {year} {2022}{\natexlab{b}})}\BibitemShut
  {NoStop}%
\bibitem [{\citenamefont {Simk{\'o}}\ \emph {et~al.}(2023)\citenamefont
  {Simk{\'o}}, \citenamefont {Schran}, \citenamefont {Brieuc}, \citenamefont
  {F{\'a}bri}, \citenamefont {Asvany}, \citenamefont {Schlemmer}, \citenamefont
  {Marx},\ and\ \citenamefont {Cs{\'a}sz{\'a}r}}]{Quantum2023simko}%
  \BibitemOpen
  \bibfield  {author} {\bibinfo {author} {\bibfnamefont {I.}~\bibnamefont
  {Simk{\'o}}}, \bibinfo {author} {\bibfnamefont {C.}~\bibnamefont {Schran}},
  \bibinfo {author} {\bibfnamefont {F.}~\bibnamefont {Brieuc}}, \bibinfo
  {author} {\bibfnamefont {C.}~\bibnamefont {F{\'a}bri}}, \bibinfo {author}
  {\bibfnamefont {O.}~\bibnamefont {Asvany}}, \bibinfo {author} {\bibfnamefont
  {S.}~\bibnamefont {Schlemmer}}, \bibinfo {author} {\bibfnamefont
  {D.}~\bibnamefont {Marx}},\ and\ \bibinfo {author} {\bibfnamefont {A.~G.}\
  \bibnamefont {Cs{\'a}sz{\'a}r}},\ }\bibfield  {title} {\enquote {\bibinfo
  {title} {Quantum {{Nuclear Delocalization}} and its {{Rovibrational
  Fingerprints}}},}\ }\href {https://doi.org/10.1002/anie.202306744} {\bibfield
   {journal} {\bibinfo  {journal} {Angew. Chem. Int. Ed.}\ }\textbf {\bibinfo
  {volume} {62}},\ \bibinfo {pages} {e202306744} (\bibinfo {year}
  {2023})}\BibitemShut {NoStop}%
\bibitem [{\citenamefont {Bowman}, \citenamefont {Carrington},\ and\
  \citenamefont {Meyer}(2008)}]{Variational2008bowman}%
  \BibitemOpen
  \bibfield  {author} {\bibinfo {author} {\bibfnamefont {J.~M.}\ \bibnamefont
  {Bowman}}, \bibinfo {author} {\bibfnamefont {T.}~\bibnamefont {Carrington}},\
  and\ \bibinfo {author} {\bibfnamefont {H.-D.}\ \bibnamefont {Meyer}},\
  }\bibfield  {title} {\enquote {\bibinfo {title} {Variational quantum
  approaches for computing vibrational energies of polyatomic molecules},}\
  }\href {https://doi.org/10.1080/00268970802258609} {\bibfield  {journal}
  {\bibinfo  {journal} {Mol. Phys.}\ }\textbf {\bibinfo {volume} {106}},\
  \bibinfo {pages} {2145--2182} (\bibinfo {year} {2008})}\BibitemShut {NoStop}%
\bibitem [{\citenamefont {Carrington}(2017)}]{Perspective2017carrington}%
  \BibitemOpen
  \bibfield  {author} {\bibinfo {author} {\bibfnamefont {T.}~\bibnamefont
  {Carrington}},\ }\bibfield  {title} {\enquote {\bibinfo {title} {Perspective:
  {{Computing}} (ro-)vibrational spectra of molecules with more than four
  atoms},}\ }\href {https://doi.org/10.1063/1.4979117} {\bibfield  {journal}
  {\bibinfo  {journal} {J. Chem. Phys.}\ }\textbf {\bibinfo {volume} {146}},\
  \bibinfo {pages} {120902} (\bibinfo {year} {2017})}\BibitemShut {NoStop}%
\bibitem [{\citenamefont {Barone}\ \emph {et~al.}(2021)\citenamefont {Barone},
  \citenamefont {Alessandrini}, \citenamefont {Biczysko}, \citenamefont
  {Cheeseman}, \citenamefont {Clary}, \citenamefont {McCoy}, \citenamefont
  {DiRisio}, \citenamefont {Neese}, \citenamefont {Melosso},\ and\
  \citenamefont {Puzzarini}}]{Computational2021barone}%
  \BibitemOpen
  \bibfield  {author} {\bibinfo {author} {\bibfnamefont {V.}~\bibnamefont
  {Barone}}, \bibinfo {author} {\bibfnamefont {S.}~\bibnamefont
  {Alessandrini}}, \bibinfo {author} {\bibfnamefont {M.}~\bibnamefont
  {Biczysko}}, \bibinfo {author} {\bibfnamefont {J.~R.}\ \bibnamefont
  {Cheeseman}}, \bibinfo {author} {\bibfnamefont {D.~C.}\ \bibnamefont
  {Clary}}, \bibinfo {author} {\bibfnamefont {A.~B.}\ \bibnamefont {McCoy}},
  \bibinfo {author} {\bibfnamefont {R.~J.}\ \bibnamefont {DiRisio}}, \bibinfo
  {author} {\bibfnamefont {F.}~\bibnamefont {Neese}}, \bibinfo {author}
  {\bibfnamefont {M.}~\bibnamefont {Melosso}},\ and\ \bibinfo {author}
  {\bibfnamefont {C.}~\bibnamefont {Puzzarini}},\ }\bibfield  {title} {\enquote
  {\bibinfo {title} {Computational molecular spectroscopy},}\ }\href
  {https://doi.org/10.1038/s43586-021-00034-1} {\bibfield  {journal} {\bibinfo
  {journal} {Nat. Rev. Methods Primer}\ }\textbf {\bibinfo {volume} {1}},\
  \bibinfo {pages} {1--27} (\bibinfo {year} {2021})}\BibitemShut {NoStop}%
\bibitem [{\citenamefont {M{\'a}tyus}, \citenamefont {Mart{\'i}n
  Santa~Dar{\'i}a},\ and\ \citenamefont {Avila}(2023)}]{Exact2023matyus}%
  \BibitemOpen
  \bibfield  {author} {\bibinfo {author} {\bibfnamefont {E.}~\bibnamefont
  {M{\'a}tyus}}, \bibinfo {author} {\bibfnamefont {A.}~\bibnamefont {Mart{\'i}n
  Santa~Dar{\'i}a}},\ and\ \bibinfo {author} {\bibfnamefont {G.}~\bibnamefont
  {Avila}},\ }\bibfield  {title} {\enquote {\bibinfo {title} {Exact quantum
  dynamics developments for floppy molecular systems and complexes},}\ }\href
  {https://doi.org/10.1039/D2CC05123K} {\bibfield  {journal} {\bibinfo
  {journal} {Chem. Commun.}\ }\textbf {\bibinfo {volume} {59}},\ \bibinfo
  {pages} {366--381} (\bibinfo {year} {2023})}\BibitemShut {NoStop}%
\bibitem [{\citenamefont {Larsson}(2024)}]{Tensor2024larsson}%
  \BibitemOpen
  \bibfield  {author} {\bibinfo {author} {\bibfnamefont {H.~R.}\ \bibnamefont
  {Larsson}},\ }\bibfield  {title} {\enquote {\bibinfo {title} {A tensor
  network view of multilayer multiconfiguration time-dependent {{Hartree}}
  methods},}\ }\href {https://doi.org/10.1080/00268976.2024.2306881} {\bibfield
   {journal} {\bibinfo  {journal} {Mol. Phys.}\ }\textbf {\bibinfo {volume}
  {122}},\ \bibinfo {pages} {e2306881} (\bibinfo {year} {2024})}\BibitemShut
  {NoStop}%
\bibitem [{\citenamefont {Goerigk}\ and\ \citenamefont
  {Grimme}(2011)}]{Thorough2011goerigk}%
  \BibitemOpen
  \bibfield  {author} {\bibinfo {author} {\bibfnamefont {L.}~\bibnamefont
  {Goerigk}}\ and\ \bibinfo {author} {\bibfnamefont {S.}~\bibnamefont
  {Grimme}},\ }\bibfield  {title} {\enquote {\bibinfo {title} {A thorough
  benchmark of density functional methods for general main group
  thermochemistry, kinetics, and noncovalent interactions},}\ }\href
  {https://doi.org/10.1039/c0cp02984j} {\bibfield  {journal} {\bibinfo
  {journal} {Phys. Chem. Chem. Phys.}\ }\textbf {\bibinfo {volume} {13}},\
  \bibinfo {pages} {6670} (\bibinfo {year} {2011})}\BibitemShut {NoStop}%
\bibitem [{\citenamefont {Motta}\ \emph {et~al.}(2017)\citenamefont {Motta},
  \citenamefont {Ceperley}, \citenamefont {Chan}, \citenamefont {Gomez},
  \citenamefont {Gull}, \citenamefont {Guo}, \citenamefont
  {{Jim{\'e}nez-Hoyos}}, \citenamefont {Lan}, \citenamefont {Li}, \citenamefont
  {Ma}, \citenamefont {Millis}, \citenamefont {Prokof'ev}, \citenamefont {Ray},
  \citenamefont {Scuseria}, \citenamefont {Sorella}, \citenamefont
  {Stoudenmire}, \citenamefont {Sun}, \citenamefont {Tupitsyn}, \citenamefont
  {White}, \citenamefont {Zgid}, \citenamefont {Zhang},\ and\ \citenamefont
  {{Simons Collaboration on the Many-Electron Problem}}}]{Solution2017motta}%
  \BibitemOpen
  \bibfield  {author} {\bibinfo {author} {\bibfnamefont {M.}~\bibnamefont
  {Motta}}, \bibinfo {author} {\bibfnamefont {D.~M.}\ \bibnamefont {Ceperley}},
  \bibinfo {author} {\bibfnamefont {G.~K.-L.}\ \bibnamefont {Chan}}, \bibinfo
  {author} {\bibfnamefont {J.~A.}\ \bibnamefont {Gomez}}, \bibinfo {author}
  {\bibfnamefont {E.}~\bibnamefont {Gull}}, \bibinfo {author} {\bibfnamefont
  {S.}~\bibnamefont {Guo}}, \bibinfo {author} {\bibfnamefont {C.~A.}\
  \bibnamefont {{Jim{\'e}nez-Hoyos}}}, \bibinfo {author} {\bibfnamefont
  {T.~N.}\ \bibnamefont {Lan}}, \bibinfo {author} {\bibfnamefont
  {J.}~\bibnamefont {Li}}, \bibinfo {author} {\bibfnamefont {F.}~\bibnamefont
  {Ma}}, \bibinfo {author} {\bibfnamefont {A.~J.}\ \bibnamefont {Millis}},
  \bibinfo {author} {\bibfnamefont {N.~V.}\ \bibnamefont {Prokof'ev}}, \bibinfo
  {author} {\bibfnamefont {U.}~\bibnamefont {Ray}}, \bibinfo {author}
  {\bibfnamefont {G.~E.}\ \bibnamefont {Scuseria}}, \bibinfo {author}
  {\bibfnamefont {S.}~\bibnamefont {Sorella}}, \bibinfo {author} {\bibfnamefont
  {E.~M.}\ \bibnamefont {Stoudenmire}}, \bibinfo {author} {\bibfnamefont
  {Q.}~\bibnamefont {Sun}}, \bibinfo {author} {\bibfnamefont {I.~S.}\
  \bibnamefont {Tupitsyn}}, \bibinfo {author} {\bibfnamefont {S.~R.}\
  \bibnamefont {White}}, \bibinfo {author} {\bibfnamefont {D.}~\bibnamefont
  {Zgid}}, \bibinfo {author} {\bibfnamefont {S.}~\bibnamefont {Zhang}},\ and\
  \bibinfo {author} {\bibnamefont {{Simons Collaboration on the Many-Electron
  Problem}}},\ }\bibfield  {title} {\enquote {\bibinfo {title} {Towards the
  {{Solution}} of the {{Many-Electron Problem}} in {{Real Materials}}:
  {{Equation}} of {{State}} of the {{Hydrogen Chain}} with {{State-of-the-Art
  Many-Body Methods}}},}\ }\href {https://doi.org/10.1103/PhysRevX.7.031059}
  {\bibfield  {journal} {\bibinfo  {journal} {Phys. Rev. X}\ }\textbf {\bibinfo
  {volume} {7}},\ \bibinfo {pages} {031059} (\bibinfo {year}
  {2017})}\BibitemShut {NoStop}%
\bibitem [{\citenamefont {Loos}\ \emph {et~al.}(2020)\citenamefont {Loos},
  \citenamefont {Lipparini}, \citenamefont {{Boggio-Pasqua}}, \citenamefont
  {Scemama},\ and\ \citenamefont {Jacquemin}}]{Mountaineering2020loos}%
  \BibitemOpen
  \bibfield  {author} {\bibinfo {author} {\bibfnamefont {P.-F.}\ \bibnamefont
  {Loos}}, \bibinfo {author} {\bibfnamefont {F.}~\bibnamefont {Lipparini}},
  \bibinfo {author} {\bibfnamefont {M.}~\bibnamefont {{Boggio-Pasqua}}},
  \bibinfo {author} {\bibfnamefont {A.}~\bibnamefont {Scemama}},\ and\ \bibinfo
  {author} {\bibfnamefont {D.}~\bibnamefont {Jacquemin}},\ }\bibfield  {title}
  {\enquote {\bibinfo {title} {A {{Mountaineering Strategy}} to {{Excited
  States}}: {{Highly Accurate Energies}} and {{Benchmarks}} for {{Medium Sized
  Molecules}}},}\ }\href {https://doi.org/10.1021/acs.jctc.9b01216} {\bibfield
  {journal} {\bibinfo  {journal} {J. Chem. Theory Comput.}\ }\textbf {\bibinfo
  {volume} {16}},\ \bibinfo {pages} {1711--1741} (\bibinfo {year}
  {2020})}\BibitemShut {NoStop}%
\bibitem [{\citenamefont {Eriksen}\ \emph {et~al.}(2020)\citenamefont
  {Eriksen}, \citenamefont {Anderson}, \citenamefont {Deustua}, \citenamefont
  {Ghanem}, \citenamefont {Hait}, \citenamefont {Hoffmann}, \citenamefont
  {Lee}, \citenamefont {Levine}, \citenamefont {Magoulas}, \citenamefont
  {Shen}, \citenamefont {Tubman}, \citenamefont {Whaley}, \citenamefont {Xu},
  \citenamefont {Yao}, \citenamefont {Zhang}, \citenamefont {Alavi},
  \citenamefont {Chan}, \citenamefont {{Head-Gordon}}, \citenamefont {Liu},
  \citenamefont {Piecuch}, \citenamefont {Sharma}, \citenamefont {{Ten-no}},
  \citenamefont {Umrigar},\ and\ \citenamefont {Gauss}}]{Ground2020eriksen}%
  \BibitemOpen
  \bibfield  {author} {\bibinfo {author} {\bibfnamefont {J.~J.}\ \bibnamefont
  {Eriksen}}, \bibinfo {author} {\bibfnamefont {T.~A.}\ \bibnamefont
  {Anderson}}, \bibinfo {author} {\bibfnamefont {J.~E.}\ \bibnamefont
  {Deustua}}, \bibinfo {author} {\bibfnamefont {K.}~\bibnamefont {Ghanem}},
  \bibinfo {author} {\bibfnamefont {D.}~\bibnamefont {Hait}}, \bibinfo {author}
  {\bibfnamefont {M.~R.}\ \bibnamefont {Hoffmann}}, \bibinfo {author}
  {\bibfnamefont {S.}~\bibnamefont {Lee}}, \bibinfo {author} {\bibfnamefont
  {D.~S.}\ \bibnamefont {Levine}}, \bibinfo {author} {\bibfnamefont
  {I.}~\bibnamefont {Magoulas}}, \bibinfo {author} {\bibfnamefont
  {J.}~\bibnamefont {Shen}}, \bibinfo {author} {\bibfnamefont {N.~M.}\
  \bibnamefont {Tubman}}, \bibinfo {author} {\bibfnamefont {K.~B.}\
  \bibnamefont {Whaley}}, \bibinfo {author} {\bibfnamefont {E.}~\bibnamefont
  {Xu}}, \bibinfo {author} {\bibfnamefont {Y.}~\bibnamefont {Yao}}, \bibinfo
  {author} {\bibfnamefont {N.}~\bibnamefont {Zhang}}, \bibinfo {author}
  {\bibfnamefont {A.}~\bibnamefont {Alavi}}, \bibinfo {author} {\bibfnamefont
  {G.~K.-L.}\ \bibnamefont {Chan}}, \bibinfo {author} {\bibfnamefont
  {M.}~\bibnamefont {{Head-Gordon}}}, \bibinfo {author} {\bibfnamefont
  {W.}~\bibnamefont {Liu}}, \bibinfo {author} {\bibfnamefont {P.}~\bibnamefont
  {Piecuch}}, \bibinfo {author} {\bibfnamefont {S.}~\bibnamefont {Sharma}},
  \bibinfo {author} {\bibfnamefont {S.~L.}\ \bibnamefont {{Ten-no}}}, \bibinfo
  {author} {\bibfnamefont {C.~J.}\ \bibnamefont {Umrigar}},\ and\ \bibinfo
  {author} {\bibfnamefont {J.}~\bibnamefont {Gauss}},\ }\bibfield  {title}
  {\enquote {\bibinfo {title} {The {{Ground State Electronic Energy}} of
  {{Benzene}}},}\ }\href {https://doi.org/10.1021/acs.jpclett.0c02621}
  {\bibfield  {journal} {\bibinfo  {journal} {J. Phys. Chem. Lett.}\ ,\
  \bibinfo {pages} {8922--8929}} (\bibinfo {year} {2020})}\BibitemShut
  {NoStop}%
\bibitem [{\citenamefont {Williams}\ \emph {et~al.}(2020)\citenamefont
  {Williams}, \citenamefont {Yao}, \citenamefont {Li}, \citenamefont {Chen},
  \citenamefont {Shi}, \citenamefont {Motta}, \citenamefont {Niu},
  \citenamefont {Ray}, \citenamefont {Guo}, \citenamefont {Anderson},
  \citenamefont {Li}, \citenamefont {Tran}, \citenamefont {Yeh}, \citenamefont
  {Mussard}, \citenamefont {Sharma}, \citenamefont {Bruneval}, \citenamefont
  {{van Schilfgaarde}}, \citenamefont {Booth}, \citenamefont {Chan},
  \citenamefont {Zhang}, \citenamefont {Gull}, \citenamefont {Zgid},
  \citenamefont {Millis}, \citenamefont {Umrigar},\ and\ \citenamefont
  {Wagner}}]{Direct2020williams}%
  \BibitemOpen
  \bibfield  {author} {\bibinfo {author} {\bibfnamefont {K.~T.}\ \bibnamefont
  {Williams}}, \bibinfo {author} {\bibfnamefont {Y.}~\bibnamefont {Yao}},
  \bibinfo {author} {\bibfnamefont {J.}~\bibnamefont {Li}}, \bibinfo {author}
  {\bibfnamefont {L.}~\bibnamefont {Chen}}, \bibinfo {author} {\bibfnamefont
  {H.}~\bibnamefont {Shi}}, \bibinfo {author} {\bibfnamefont {M.}~\bibnamefont
  {Motta}}, \bibinfo {author} {\bibfnamefont {C.}~\bibnamefont {Niu}}, \bibinfo
  {author} {\bibfnamefont {U.}~\bibnamefont {Ray}}, \bibinfo {author}
  {\bibfnamefont {S.}~\bibnamefont {Guo}}, \bibinfo {author} {\bibfnamefont
  {R.~J.}\ \bibnamefont {Anderson}}, \bibinfo {author} {\bibfnamefont
  {J.}~\bibnamefont {Li}}, \bibinfo {author} {\bibfnamefont {L.~N.}\
  \bibnamefont {Tran}}, \bibinfo {author} {\bibfnamefont {C.-N.}\ \bibnamefont
  {Yeh}}, \bibinfo {author} {\bibfnamefont {B.}~\bibnamefont {Mussard}},
  \bibinfo {author} {\bibfnamefont {S.}~\bibnamefont {Sharma}}, \bibinfo
  {author} {\bibfnamefont {F.}~\bibnamefont {Bruneval}}, \bibinfo {author}
  {\bibfnamefont {M.}~\bibnamefont {{van Schilfgaarde}}}, \bibinfo {author}
  {\bibfnamefont {G.~H.}\ \bibnamefont {Booth}}, \bibinfo {author}
  {\bibfnamefont {G.~K.-L.}\ \bibnamefont {Chan}}, \bibinfo {author}
  {\bibfnamefont {S.}~\bibnamefont {Zhang}}, \bibinfo {author} {\bibfnamefont
  {E.}~\bibnamefont {Gull}}, \bibinfo {author} {\bibfnamefont {D.}~\bibnamefont
  {Zgid}}, \bibinfo {author} {\bibfnamefont {A.}~\bibnamefont {Millis}},
  \bibinfo {author} {\bibfnamefont {C.~J.}\ \bibnamefont {Umrigar}},\ and\
  \bibinfo {author} {\bibfnamefont {L.~K.}\ \bibnamefont {Wagner}},\ }\bibfield
   {title} {\enquote {\bibinfo {title} {Direct {{Comparison}} of {{Many-Body
  Methods}} for {{Realistic Electronic Hamiltonians}}},}\ }\href
  {https://doi.org/10.1103/PhysRevX.10.011041} {\bibfield  {journal} {\bibinfo
  {journal} {Phys. Rev. X}\ }\textbf {\bibinfo {volume} {10}},\ \bibinfo
  {pages} {011041} (\bibinfo {year} {2020})}\BibitemShut {NoStop}%
\bibitem [{\citenamefont {Larsson}, \citenamefont {{Jim{\'e}nez-Hoyos}},\ and\
  \citenamefont {Chan}(2020)}]{Minimal2020larsson}%
  \BibitemOpen
  \bibfield  {author} {\bibinfo {author} {\bibfnamefont {H.~R.}\ \bibnamefont
  {Larsson}}, \bibinfo {author} {\bibfnamefont {C.~A.}\ \bibnamefont
  {{Jim{\'e}nez-Hoyos}}},\ and\ \bibinfo {author} {\bibfnamefont {G.~K.-L.}\
  \bibnamefont {Chan}},\ }\bibfield  {title} {\enquote {\bibinfo {title}
  {Minimal {{Matrix Product States}} and {{Generalizations}} of {{Mean-Field}}
  and {{Geminal Wave Functions}}},}\ }\href
  {https://doi.org/10.1021/acs.jctc.0c00463} {\bibfield  {journal} {\bibinfo
  {journal} {J. Chem. Theory Comput.}\ }\textbf {\bibinfo {volume} {16}},\
  \bibinfo {pages} {5057--5066} (\bibinfo {year} {2020})}\BibitemShut {NoStop}%
\bibitem [{\citenamefont {Mata}\ and\ \citenamefont
  {Suhm}(2017)}]{Benchmarking2017mata}%
  \BibitemOpen
  \bibfield  {author} {\bibinfo {author} {\bibfnamefont {R.~A.}\ \bibnamefont
  {Mata}}\ and\ \bibinfo {author} {\bibfnamefont {M.~A.}\ \bibnamefont
  {Suhm}},\ }\bibfield  {title} {\enquote {\bibinfo {title} {Benchmarking
  {{Quantum Chemical Methods}}: {{Are We Heading}} in the {{Right
  Direction}}?}}\ }\href {https://doi.org/10.1002/anie.201611308} {\bibfield
  {journal} {\bibinfo  {journal} {Angew. Chem. Int. Ed.}\ }\textbf {\bibinfo
  {volume} {56}},\ \bibinfo {pages} {11011--11018} (\bibinfo {year}
  {2017})}\BibitemShut {NoStop}%
\bibitem [{\citenamefont {Welsch}(2018)}]{LowTemperature2018welsch}%
  \BibitemOpen
  \bibfield  {author} {\bibinfo {author} {\bibfnamefont {R.}~\bibnamefont
  {Welsch}},\ }\bibfield  {title} {\enquote {\bibinfo {title}
  {Low-{{Temperature Thermal Rate Constants}} for the {{Water Formation
  Reaction H}}{\textsubscript{2}}+{{OH}} from {{Rigorous Quantum Dynamics
  Calculations}}},}\ }\href {https://doi.org/10.1002/ange.201807666} {\bibfield
   {journal} {\bibinfo  {journal} {Angew. Chem.}\ }\textbf {\bibinfo {volume}
  {130}},\ \bibinfo {pages} {13334--13337} (\bibinfo {year}
  {2018})}\BibitemShut {NoStop}%
\bibitem [{\citenamefont {A.~Mata}, \citenamefont {{Zehnacker-Rentien}},\ and\
  \citenamefont {A.~Suhm}(2023)}]{Benchmark2023a.mata}%
  \BibitemOpen
  \bibfield  {author} {\bibinfo {author} {\bibfnamefont {R.}~\bibnamefont
  {A.~Mata}}, \bibinfo {author} {\bibfnamefont {A.}~\bibnamefont
  {{Zehnacker-Rentien}}},\ and\ \bibinfo {author} {\bibfnamefont
  {M.}~\bibnamefont {A.~Suhm}},\ }\bibfield  {title} {\enquote {\bibinfo
  {title} {Benchmark experiments for numerical quantum chemistry},}\ }\href
  {https://doi.org/10.1039/D3CP90186F} {\bibfield  {journal} {\bibinfo
  {journal} {Phys. Chem. Chem. Phys.}\ }\textbf {\bibinfo {volume} {25}},\
  \bibinfo {pages} {26415--26416} (\bibinfo {year} {2023})}\BibitemShut
  {NoStop}%
\bibitem [{\citenamefont {Carbonniere}, \citenamefont {B{\'e}gu{\'e}},\ and\
  \citenamefont {Pouchan}(2004)}]{DFT2004carbonniere}%
  \BibitemOpen
  \bibfield  {author} {\bibinfo {author} {\bibfnamefont {P.}~\bibnamefont
  {Carbonniere}}, \bibinfo {author} {\bibfnamefont {D.}~\bibnamefont
  {B{\'e}gu{\'e}}},\ and\ \bibinfo {author} {\bibfnamefont {C.}~\bibnamefont
  {Pouchan}},\ }\bibfield  {title} {\enquote {\bibinfo {title} {{{DFT}} quartic
  force field of acetonitrile by using a generalized least-squares
  procedure},}\ }\href {https://doi.org/10.1016/j.cplett.2004.05.109}
  {\bibfield  {journal} {\bibinfo  {journal} {Chem. Phys. Lett.}\ }\textbf
  {\bibinfo {volume} {393}},\ \bibinfo {pages} {92--97} (\bibinfo {year}
  {2004})}\BibitemShut {NoStop}%
\bibitem [{\citenamefont {Begue}, \citenamefont {Carbonniere},\ and\
  \citenamefont {Pouchan}(2005)}]{Calculations2005begue}%
  \BibitemOpen
  \bibfield  {author} {\bibinfo {author} {\bibfnamefont {D.}~\bibnamefont
  {Begue}}, \bibinfo {author} {\bibfnamefont {P.}~\bibnamefont {Carbonniere}},\
  and\ \bibinfo {author} {\bibfnamefont {C.}~\bibnamefont {Pouchan}},\
  }\bibfield  {title} {\enquote {\bibinfo {title} {Calculations of
  {{Vibrational Energy Levels}} by {{Using}} a {{Hybrid}} ab {{Initio}} and
  {{DFT Quartic Force Field}}: {{Application}} to {{Acetonitrile}}},}\ }\href
  {https://doi.org/10.1021/jp0406114} {\bibfield  {journal} {\bibinfo
  {journal} {J. Phys. Chem. A}\ }\textbf {\bibinfo {volume} {109}},\ \bibinfo
  {pages} {4611--4616} (\bibinfo {year} {2005})}\BibitemShut {NoStop}%
\bibitem [{\citenamefont {Avila}\ and\ \citenamefont
  {Carrington}(2011{\natexlab{a}})}]{Nonproduct2011avila}%
  \BibitemOpen
  \bibfield  {author} {\bibinfo {author} {\bibfnamefont {G.}~\bibnamefont
  {Avila}}\ and\ \bibinfo {author} {\bibfnamefont {T.}~\bibnamefont
  {Carrington}},\ }\bibfield  {title} {\enquote {\bibinfo {title} {Nonproduct
  {{Quadrature Grids}}: {{Solving}} the {{Vibrational Schr{\"o}dinger
  Equation}} in 12d},}\ }in\ \href
  {https://doi.org/10.1007/978-1-4419-9491-2_1} {\emph {\bibinfo {booktitle}
  {Quantum {{Dynamic Imaging}}}}},\ \bibinfo {editor} {edited by\ \bibinfo
  {editor} {\bibfnamefont {A.~D.}\ \bibnamefont {Bandrauk}}\ and\ \bibinfo
  {editor} {\bibfnamefont {M.}~\bibnamefont {Ivanov}}}\ (\bibinfo  {publisher}
  {Springer New York},\ \bibinfo {address} {New York, NY},\ \bibinfo {year}
  {2011})\ pp.\ \bibinfo {pages} {1--12}\BibitemShut {NoStop}%
\bibitem [{\citenamefont {Avila}\ and\ \citenamefont
  {Carrington}(2011{\natexlab{b}})}]{Using2011avila}%
  \BibitemOpen
  \bibfield  {author} {\bibinfo {author} {\bibfnamefont {G.}~\bibnamefont
  {Avila}}\ and\ \bibinfo {author} {\bibfnamefont {T.}~\bibnamefont
  {Carrington}},\ }\bibfield  {title} {\enquote {\bibinfo {title} {Using
  nonproduct quadrature grids to solve the vibrational {{Schr{\"o}dinger}}
  equation in {{12D}}},}\ }\href {https://doi.org/10.1063/1.3549817} {\bibfield
   {journal} {\bibinfo  {journal} {J. Chem. Phys.}\ }\textbf {\bibinfo {volume}
  {134}},\ \bibinfo {pages} {054126} (\bibinfo {year}
  {2011}{\natexlab{b}})}\BibitemShut {NoStop}%
\bibitem [{\citenamefont {Leclerc}\ and\ \citenamefont
  {Carrington}(2014)}]{Calculating2014leclerc}%
  \BibitemOpen
  \bibfield  {author} {\bibinfo {author} {\bibfnamefont {A.}~\bibnamefont
  {Leclerc}}\ and\ \bibinfo {author} {\bibfnamefont {T.}~\bibnamefont
  {Carrington}},\ }\bibfield  {title} {\enquote {\bibinfo {title} {Calculating
  vibrational spectra with sum of product basis functions without storing
  full-dimensional vectors or matrices},}\ }\href
  {https://doi.org/10.1063/1.4871981} {\bibfield  {journal} {\bibinfo
  {journal} {J. Chem. Phys.}\ }\textbf {\bibinfo {volume} {140}},\ \bibinfo
  {pages} {174111} (\bibinfo {year} {2014})}\BibitemShut {NoStop}%
\bibitem [{\citenamefont {Lutz}\ \emph {et~al.}(2014)\citenamefont {Lutz},
  \citenamefont {Rode}, \citenamefont {Bonn},\ and\ \citenamefont
  {Huck}}]{Impact2014lutz}%
  \BibitemOpen
  \bibfield  {author} {\bibinfo {author} {\bibfnamefont {O.~M.~D.}\
  \bibnamefont {Lutz}}, \bibinfo {author} {\bibfnamefont {B.~M.}\ \bibnamefont
  {Rode}}, \bibinfo {author} {\bibfnamefont {G.~K.}\ \bibnamefont {Bonn}},\
  and\ \bibinfo {author} {\bibfnamefont {C.~W.}\ \bibnamefont {Huck}},\
  }\bibfield  {title} {\enquote {\bibinfo {title} {The impact of highly
  correlated potential energy surfaces on the anharmonically corrected {{IR}}
  spectrum of acetonitrile},}\ }\href
  {https://doi.org/10.1016/j.saa.2014.04.067} {\bibfield  {journal} {\bibinfo
  {journal} {Spectrochim. Acta. A. Mol. Biomol. Spectrosc.}\ }\textbf {\bibinfo
  {volume} {131}},\ \bibinfo {pages} {545--555} (\bibinfo {year}
  {2014})}\BibitemShut {NoStop}%
\bibitem [{\citenamefont {Halverson}\ and\ \citenamefont
  {Poirier}(2015)}]{Large2015halverson}%
  \BibitemOpen
  \bibfield  {author} {\bibinfo {author} {\bibfnamefont {T.}~\bibnamefont
  {Halverson}}\ and\ \bibinfo {author} {\bibfnamefont {B.}~\bibnamefont
  {Poirier}},\ }\bibfield  {title} {\enquote {\bibinfo {title} {Large scale
  exact quantum dynamics calculations: {{Ten}} thousand quantum states of
  acetonitrile},}\ }\href {https://doi.org/10.1016/j.cplett.2015.02.004}
  {\bibfield  {journal} {\bibinfo  {journal} {Chem. Phys. Lett.}\ }\textbf
  {\bibinfo {volume} {624}},\ \bibinfo {pages} {37--42} (\bibinfo {year}
  {2015})}\BibitemShut {NoStop}%
\bibitem [{\citenamefont {Thomas}\ and\ \citenamefont
  {Carrington}(2015)}]{Using2015thomas}%
  \BibitemOpen
  \bibfield  {author} {\bibinfo {author} {\bibfnamefont {P.~S.}\ \bibnamefont
  {Thomas}}\ and\ \bibinfo {author} {\bibfnamefont {T.}~\bibnamefont
  {Carrington}},\ }\bibfield  {title} {\enquote {\bibinfo {title} {Using
  {{Nested Contractions}} and a {{Hierarchical Tensor Format To Compute
  Vibrational Spectra}} of {{Molecules}} with {{Seven Atoms}}},}\ }\href
  {https://doi.org/10.1021/acs.jpca.5b10015} {\bibfield  {journal} {\bibinfo
  {journal} {J. Phys. Chem. A}\ }\textbf {\bibinfo {volume} {119}},\ \bibinfo
  {pages} {13074--13091} (\bibinfo {year} {2015})}\BibitemShut {NoStop}%
\bibitem [{\citenamefont {Brown}\ and\ \citenamefont
  {Carrington}(2016)}]{Using2016browna}%
  \BibitemOpen
  \bibfield  {author} {\bibinfo {author} {\bibfnamefont {J.}~\bibnamefont
  {Brown}}\ and\ \bibinfo {author} {\bibfnamefont {T.}~\bibnamefont
  {Carrington}},\ }\bibfield  {title} {\enquote {\bibinfo {title} {Using an
  expanding nondirect product harmonic basis with an iterative eigensolver to
  compute vibrational energy levels with as many as seven atoms},}\ }\href
  {https://doi.org/10.1063/1.4963916} {\bibfield  {journal} {\bibinfo
  {journal} {J. Chem. Phys.}\ }\textbf {\bibinfo {volume} {145}},\ \bibinfo
  {pages} {144104} (\bibinfo {year} {2016})}\BibitemShut {NoStop}%
\bibitem [{\citenamefont {Garnier}\ \emph {et~al.}(2016)\citenamefont
  {Garnier}, \citenamefont {Odunlami}, \citenamefont {Le~Bris}, \citenamefont
  {B{\'e}gu{\'e}}, \citenamefont {Baraille},\ and\ \citenamefont
  {Coulaud}}]{Adaptive2016garnier}%
  \BibitemOpen
  \bibfield  {author} {\bibinfo {author} {\bibfnamefont {R.}~\bibnamefont
  {Garnier}}, \bibinfo {author} {\bibfnamefont {M.}~\bibnamefont {Odunlami}},
  \bibinfo {author} {\bibfnamefont {V.}~\bibnamefont {Le~Bris}}, \bibinfo
  {author} {\bibfnamefont {D.}~\bibnamefont {B{\'e}gu{\'e}}}, \bibinfo {author}
  {\bibfnamefont {I.}~\bibnamefont {Baraille}},\ and\ \bibinfo {author}
  {\bibfnamefont {O.}~\bibnamefont {Coulaud}},\ }\bibfield  {title} {\enquote
  {\bibinfo {title} {Adaptive vibrational configuration interaction
  ({{A-VCI}}): {{{\emph{A}}}}{\emph{ posteriori}} error estimation to
  efficiently compute anharmonic {{IR}} spectra},}\ }\href
  {https://doi.org/10.1063/1.4952414} {\bibfield  {journal} {\bibinfo
  {journal} {J. Chem. Phys.}\ }\textbf {\bibinfo {volume} {144}},\ \bibinfo
  {pages} {204123} (\bibinfo {year} {2016})}\BibitemShut {NoStop}%
\bibitem [{\citenamefont {Leclerc}\ and\ \citenamefont
  {Carrington}(2016)}]{Using2016leclerc}%
  \BibitemOpen
  \bibfield  {author} {\bibinfo {author} {\bibfnamefont {A.}~\bibnamefont
  {Leclerc}}\ and\ \bibinfo {author} {\bibfnamefont {T.}~\bibnamefont
  {Carrington}},\ }\bibfield  {title} {\enquote {\bibinfo {title} {Using
  symmetry-adapted optimized sum-of-products basis functions to calculate
  vibrational spectra},}\ }\href {https://doi.org/10.1016/j.cplett.2015.11.057}
  {\bibfield  {journal} {\bibinfo  {journal} {Chem. Phys. Lett.}\ }\textbf
  {\bibinfo {volume} {644}},\ \bibinfo {pages} {183--188} (\bibinfo {year}
  {2016})}\BibitemShut {NoStop}%
\bibitem [{\citenamefont {Leclerc}, \citenamefont {Thomas},\ and\ \citenamefont
  {Carrington}(2016)}]{Comparison2016leclerc}%
  \BibitemOpen
  \bibfield  {author} {\bibinfo {author} {\bibfnamefont {A.}~\bibnamefont
  {Leclerc}}, \bibinfo {author} {\bibfnamefont {P.~S.}\ \bibnamefont
  {Thomas}},\ and\ \bibinfo {author} {\bibfnamefont {T.}~\bibnamefont
  {Carrington}},\ }\bibfield  {title} {\enquote {\bibinfo {title} {Comparison
  of different eigensolvers for calculating vibrational spectra using low-rank,
  sum-of-product basis functions},}\ }\href
  {https://doi.org/10.1080/00268976.2016.1249980} {\bibfield  {journal}
  {\bibinfo  {journal} {Mol. Phys.}\ ,\ \bibinfo {pages} {1--10}} (\bibinfo
  {year} {2016})}\BibitemShut {NoStop}%
\bibitem [{\citenamefont {Rakhuba}\ and\ \citenamefont
  {Oseledets}(2016)}]{Calculating2016rakhuba}%
  \BibitemOpen
  \bibfield  {author} {\bibinfo {author} {\bibfnamefont {M.}~\bibnamefont
  {Rakhuba}}\ and\ \bibinfo {author} {\bibfnamefont {I.}~\bibnamefont
  {Oseledets}},\ }\bibfield  {title} {\enquote {\bibinfo {title} {Calculating
  vibrational spectra of molecules using tensor train decomposition},}\ }\href
  {https://doi.org/10.1063/1.4962420} {\bibfield  {journal} {\bibinfo
  {journal} {J. Chem. Phys.}\ }\textbf {\bibinfo {volume} {145}},\ \bibinfo
  {pages} {124101} (\bibinfo {year} {2016})}\BibitemShut {NoStop}%
\bibitem [{\citenamefont {Wodraszka}\ and\ \citenamefont
  {Carrington}(2016)}]{Using2016wodraszka}%
  \BibitemOpen
  \bibfield  {author} {\bibinfo {author} {\bibfnamefont {R.}~\bibnamefont
  {Wodraszka}}\ and\ \bibinfo {author} {\bibfnamefont {T.}~\bibnamefont
  {Carrington}},\ }\bibfield  {title} {\enquote {\bibinfo {title} {Using a
  pruned, nondirect product basis in conjunction with the multi-configuration
  time-dependent {{Hartree}} ({{MCTDH}}) method},}\ }\href
  {https://doi.org/10.1063/1.4959228} {\bibfield  {journal} {\bibinfo
  {journal} {J. Chem. Phys.}\ }\textbf {\bibinfo {volume} {145}},\ \bibinfo
  {pages} {044110} (\bibinfo {year} {2016})}\BibitemShut {NoStop}%
\bibitem [{\citenamefont {Avila}\ and\ \citenamefont
  {Carrington}(2017)}]{Pruned2017avila}%
  \BibitemOpen
  \bibfield  {author} {\bibinfo {author} {\bibfnamefont {G.}~\bibnamefont
  {Avila}}\ and\ \bibinfo {author} {\bibfnamefont {T.}~\bibnamefont
  {Carrington}},\ }\bibfield  {title} {\enquote {\bibinfo {title} {Pruned bases
  that are compatible with iterative eigensolvers and general potentials:
  {{New}} results for {{CH}}{\textsubscript{3}}{{CN}}},}\ }\href
  {https://doi.org/10.1016/j.chemphys.2016.09.023} {\bibfield  {journal}
  {\bibinfo  {journal} {Chem. Phys.}\ }\textbf {\bibinfo {volume} {482}},\
  \bibinfo {pages} {3--8} (\bibinfo {year} {2017})}\BibitemShut {NoStop}%
\bibitem [{\citenamefont {Baiardi}\ \emph {et~al.}(2017)\citenamefont
  {Baiardi}, \citenamefont {Stein}, \citenamefont {Barone},\ and\ \citenamefont
  {Reiher}}]{Vibrational2017baiardi}%
  \BibitemOpen
  \bibfield  {author} {\bibinfo {author} {\bibfnamefont {A.}~\bibnamefont
  {Baiardi}}, \bibinfo {author} {\bibfnamefont {C.~J.}\ \bibnamefont {Stein}},
  \bibinfo {author} {\bibfnamefont {V.}~\bibnamefont {Barone}},\ and\ \bibinfo
  {author} {\bibfnamefont {M.}~\bibnamefont {Reiher}},\ }\bibfield  {title}
  {\enquote {\bibinfo {title} {Vibrational {{Density Matrix Renormalization
  Group}}},}\ }\href {https://doi.org/10.1021/acs.jctc.7b00329} {\bibfield
  {journal} {\bibinfo  {journal} {J. Chem. Theory Comput.}\ }\textbf {\bibinfo
  {volume} {13}},\ \bibinfo {pages} {3764--3777} (\bibinfo {year}
  {2017})}\BibitemShut {NoStop}%
\bibitem [{\citenamefont {Odunlami}\ \emph {et~al.}(2017)\citenamefont
  {Odunlami}, \citenamefont {Le~Bris}, \citenamefont {B{\'e}gu{\'e}},
  \citenamefont {Baraille},\ and\ \citenamefont {Coulaud}}]{AVCI2017odunlami}%
  \BibitemOpen
  \bibfield  {author} {\bibinfo {author} {\bibfnamefont {M.}~\bibnamefont
  {Odunlami}}, \bibinfo {author} {\bibfnamefont {V.}~\bibnamefont {Le~Bris}},
  \bibinfo {author} {\bibfnamefont {D.}~\bibnamefont {B{\'e}gu{\'e}}}, \bibinfo
  {author} {\bibfnamefont {I.}~\bibnamefont {Baraille}},\ and\ \bibinfo
  {author} {\bibfnamefont {O.}~\bibnamefont {Coulaud}},\ }\bibfield  {title}
  {\enquote {\bibinfo {title} {A-{{VCI}}: {{A}} flexible method to efficiently
  compute vibrational spectra},}\ }\href {https://doi.org/10.1063/1.4984266}
  {\bibfield  {journal} {\bibinfo  {journal} {J. Chem. Phys.}\ }\textbf
  {\bibinfo {volume} {146}},\ \bibinfo {pages} {214108} (\bibinfo {year}
  {2017})}\BibitemShut {NoStop}%
\bibitem [{\citenamefont {Wodraszka}\ and\ \citenamefont
  {Carrington}(2017)}]{Systematically2017wodraszka}%
  \BibitemOpen
  \bibfield  {author} {\bibinfo {author} {\bibfnamefont {R.}~\bibnamefont
  {Wodraszka}}\ and\ \bibinfo {author} {\bibfnamefont {T.}~\bibnamefont
  {Carrington}},\ }\bibfield  {title} {\enquote {\bibinfo {title}
  {Systematically expanding nondirect product bases within the pruned
  multi-configuration time-dependent {{Hartree}} ({{MCTDH}}) method: {{A}}
  comparison with multi-layer {{MCTDH}}},}\ }\href
  {https://doi.org/10.1063/1.4983281} {\bibfield  {journal} {\bibinfo
  {journal} {J. Chem. Phys.}\ }\textbf {\bibinfo {volume} {146}},\ \bibinfo
  {pages} {194105} (\bibinfo {year} {2017})}\BibitemShut {NoStop}%
\bibitem [{\citenamefont {Larsson}(2019)}]{Computing2019larsson}%
  \BibitemOpen
  \bibfield  {author} {\bibinfo {author} {\bibfnamefont {H.~R.}\ \bibnamefont
  {Larsson}},\ }\bibfield  {title} {\enquote {\bibinfo {title} {Computing
  vibrational eigenstates with tree tensor network states ({{TTNS}})},}\ }\href
  {https://doi.org/10.1063/1.5130390} {\bibfield  {journal} {\bibinfo
  {journal} {J. Chem. Phys.}\ }\textbf {\bibinfo {volume} {151}},\ \bibinfo
  {pages} {204102} (\bibinfo {year} {2019})}\BibitemShut {NoStop}%
\bibitem [{\citenamefont {Garnier}(2019)}]{Dual2019garnier}%
  \BibitemOpen
  \bibfield  {author} {\bibinfo {author} {\bibfnamefont {R.}~\bibnamefont
  {Garnier}},\ }\bibfield  {title} {\enquote {\bibinfo {title} {Dual vibration
  configuration interaction ({{DVCI}}). {{An}} efficient factorization of
  molecular {{Hamiltonian}} for high performance infrared spectrum
  computation},}\ }\href {https://doi.org/10.1016/j.cpc.2018.07.008} {\bibfield
   {journal} {\bibinfo  {journal} {Comput. Phys. Commun.}\ }\textbf {\bibinfo
  {volume} {234}},\ \bibinfo {pages} {263--277} (\bibinfo {year}
  {2019})}\BibitemShut {NoStop}%
\bibitem [{\citenamefont {Lesko}, \citenamefont {Ardiansyah},\ and\
  \citenamefont {Brorsen}(2019)}]{Vibrational2019lesko}%
  \BibitemOpen
  \bibfield  {author} {\bibinfo {author} {\bibfnamefont {E.}~\bibnamefont
  {Lesko}}, \bibinfo {author} {\bibfnamefont {M.}~\bibnamefont {Ardiansyah}},\
  and\ \bibinfo {author} {\bibfnamefont {K.~R.}\ \bibnamefont {Brorsen}},\
  }\bibfield  {title} {\enquote {\bibinfo {title} {Vibrational adaptive
  sampling configuration interaction},}\ }\href
  {https://doi.org/10.1063/1.5126510} {\bibfield  {journal} {\bibinfo
  {journal} {J. Chem. Phys.}\ }\textbf {\bibinfo {volume} {151}},\ \bibinfo
  {pages} {164103} (\bibinfo {year} {2019})}\BibitemShut {NoStop}%
\bibitem [{\citenamefont {Rakhuba}, \citenamefont {Novikov},\ and\
  \citenamefont {Oseledets}(2019)}]{Lowrank2019rakhuba}%
  \BibitemOpen
  \bibfield  {author} {\bibinfo {author} {\bibfnamefont {M.}~\bibnamefont
  {Rakhuba}}, \bibinfo {author} {\bibfnamefont {A.}~\bibnamefont {Novikov}},\
  and\ \bibinfo {author} {\bibfnamefont {I.}~\bibnamefont {Oseledets}},\
  }\bibfield  {title} {\enquote {\bibinfo {title} {Low-rank {{Riemannian}}
  eigensolver for high-dimensional {{Hamiltonians}}},}\ }\href
  {https://doi.org/10.1016/j.jcp.2019.07.003} {\bibfield  {journal} {\bibinfo
  {journal} {J. Comput. Phys.}\ }\textbf {\bibinfo {volume} {396}},\ \bibinfo
  {pages} {718--737} (\bibinfo {year} {2019})}\BibitemShut {NoStop}%
\bibitem [{\citenamefont {Wodraszka}\ and\ \citenamefont
  {Carrington}(2020)}]{Collocationbased2020wodraszka}%
  \BibitemOpen
  \bibfield  {author} {\bibinfo {author} {\bibfnamefont {R.}~\bibnamefont
  {Wodraszka}}\ and\ \bibinfo {author} {\bibfnamefont {T.}~\bibnamefont
  {Carrington}},\ }\bibfield  {title} {\enquote {\bibinfo {title} {A
  collocation-based multi-configuration time-dependent {{Hartree}} method using
  mode combination and improved relaxation},}\ }\href
  {https://doi.org/10.1063/5.0006081} {\bibfield  {journal} {\bibinfo
  {journal} {J. Chem. Phys.}\ }\textbf {\bibinfo {volume} {152}},\ \bibinfo
  {pages} {164117} (\bibinfo {year} {2020})}\BibitemShut {NoStop}%
\bibitem [{\citenamefont {Fetherolf}\ and\ \citenamefont
  {Berkelbach}(2021)}]{Vibrational2021fetherolf}%
  \BibitemOpen
  \bibfield  {author} {\bibinfo {author} {\bibfnamefont {J.~H.}\ \bibnamefont
  {Fetherolf}}\ and\ \bibinfo {author} {\bibfnamefont {T.~C.}\ \bibnamefont
  {Berkelbach}},\ }\bibfield  {title} {\enquote {\bibinfo {title} {Vibrational
  heat-bath configuration interaction},}\ }\href
  {https://doi.org/10.1063/5.0035454} {\bibfield  {journal} {\bibinfo
  {journal} {J. Chem. Phys.}\ }\textbf {\bibinfo {volume} {154}},\ \bibinfo
  {pages} {074104} (\bibinfo {year} {2021})}\BibitemShut {NoStop}%
\bibitem [{\citenamefont {Kallullathil}\ and\ \citenamefont
  {Carrington}(2021)}]{Computing2021kallullathil}%
  \BibitemOpen
  \bibfield  {author} {\bibinfo {author} {\bibfnamefont {S.~D.}\ \bibnamefont
  {Kallullathil}}\ and\ \bibinfo {author} {\bibfnamefont {T.}~\bibnamefont
  {Carrington}},\ }\bibfield  {title} {\enquote {\bibinfo {title} {Computing
  vibrational energy levels by solving linear equations using a tensor method
  with an imposed rank},}\ }\href {https://doi.org/10.1063/5.0075412}
  {\bibfield  {journal} {\bibinfo  {journal} {J. Chem. Phys.}\ }\textbf
  {\bibinfo {volume} {155}},\ \bibinfo {pages} {234105} (\bibinfo {year}
  {2021})}\BibitemShut {NoStop}%
\bibitem [{\citenamefont {Sarka}\ and\ \citenamefont
  {Poirier}(2021)}]{Hitting2021sarka}%
  \BibitemOpen
  \bibfield  {author} {\bibinfo {author} {\bibfnamefont {J.}~\bibnamefont
  {Sarka}}\ and\ \bibinfo {author} {\bibfnamefont {B.}~\bibnamefont
  {Poirier}},\ }\bibfield  {title} {\enquote {\bibinfo {title} {Hitting the
  {{Trifecta}}: {{How}} to {{Simultaneously Push}} the {{Limits}} of
  {{Schr{\"o}dinger Solution}} with {{Respect}} to {{System Size}},
  {{Convergence Accuracy}}, and {{Number}} of {{Computed States}}},}\ }\href
  {https://doi.org/10.1021/acs.jctc.1c00824} {\bibfield  {journal} {\bibinfo
  {journal} {J. Chem. Theory Comput.}\ }\textbf {\bibinfo {volume} {17}},\
  \bibinfo {pages} {7732--7744} (\bibinfo {year} {2021})}\BibitemShut {NoStop}%
\bibitem [{\citenamefont {Kallullathil}\ and\ \citenamefont
  {Carrington.}(2023)}]{Computing2023kallullathil}%
  \BibitemOpen
  \bibfield  {author} {\bibinfo {author} {\bibfnamefont {S.~D.}\ \bibnamefont
  {Kallullathil}}\ and\ \bibinfo {author} {\bibfnamefont {T.}~\bibnamefont
  {Carrington.}, \bibfnamefont {Jr.}},\ }\bibfield  {title} {\enquote {\bibinfo
  {title} {Computing vibrational energy levels using a canonical polyadic
  tensor method with a fixed rank and a contraction tree},}\ }\href
  {https://doi.org/10.1063/5.0149832} {\bibfield  {journal} {\bibinfo
  {journal} {J. Chem. Phys.}\ }\textbf {\bibinfo {volume} {158}},\ \bibinfo
  {pages} {214102} (\bibinfo {year} {2023})}\BibitemShut {NoStop}%
\bibitem [{\citenamefont {Simmons}\ and\ \citenamefont
  {Carrington}(2023)}]{Computing2023simmonsa}%
  \BibitemOpen
  \bibfield  {author} {\bibinfo {author} {\bibfnamefont {J.}~\bibnamefont
  {Simmons}}\ and\ \bibinfo {author} {\bibfnamefont {T.}~\bibnamefont
  {Carrington}},\ }\bibfield  {title} {\enquote {\bibinfo {title} {Computing
  vibrational spectra using a new collocation method with a pruned basis and
  more points than basis functions: {{Avoiding}} quadrature},}\ }\href
  {https://doi.org/10.1063/5.0146703} {\bibfield  {journal} {\bibinfo
  {journal} {J. Chem. Phys.}\ }\textbf {\bibinfo {volume} {158}},\ \bibinfo
  {pages} {144115} (\bibinfo {year} {2023})}\BibitemShut {NoStop}%
\bibitem [{\citenamefont {Tran}\ and\ \citenamefont
  {Berkelbach}(2023)}]{Vibrational2023trana}%
  \BibitemOpen
  \bibfield  {author} {\bibinfo {author} {\bibfnamefont {H.~K.}\ \bibnamefont
  {Tran}}\ and\ \bibinfo {author} {\bibfnamefont {T.~C.}\ \bibnamefont
  {Berkelbach}},\ }\bibfield  {title} {\enquote {\bibinfo {title} {Vibrational
  heat-bath configuration interaction with semistochastic perturbation theory
  using harmonic oscillator or {{VSCF}} modals},}\ }\href
  {https://doi.org/10.1063/5.0172702} {\bibfield  {journal} {\bibinfo
  {journal} {J. Chem. Phys.}\ }\textbf {\bibinfo {volume} {159}},\ \bibinfo
  {pages} {194101} (\bibinfo {year} {2023})}\BibitemShut {NoStop}%
\bibitem [{\citenamefont {Hoppe}\ and\ \citenamefont
  {Manthe}(2024)}]{Eigenstate2024hoppe}%
  \BibitemOpen
  \bibfield  {author} {\bibinfo {author} {\bibfnamefont {H.}~\bibnamefont
  {Hoppe}}\ and\ \bibinfo {author} {\bibfnamefont {U.}~\bibnamefont {Manthe}},\
  }\bibfield  {title} {\enquote {\bibinfo {title} {Eigenstate calculation in
  the state-averaged (multi-layer) multi-configurational time-dependent
  {{Hartree}} approach},}\ }\href {https://doi.org/10.1063/5.0188748}
  {\bibfield  {journal} {\bibinfo  {journal} {J. Chem. Phys.}\ }\textbf
  {\bibinfo {volume} {160}},\ \bibinfo {pages} {034104} (\bibinfo {year}
  {2024})}\BibitemShut {NoStop}%
\bibitem [{\citenamefont {Rey}\ and\ \citenamefont
  {Carrington}(2024)}]{Using2024rey}%
  \BibitemOpen
  \bibfield  {author} {\bibinfo {author} {\bibfnamefont {M.}~\bibnamefont
  {Rey}}\ and\ \bibinfo {author} {\bibfnamefont {T.}~\bibnamefont {Carrington},
  \bibfnamefont {Jr.}},\ }\bibfield  {title} {\enquote {\bibinfo {title} {Using
  nested tensor train contracted basis functions with group theoretical
  techniques to compute (ro)-vibrational spectra of molecules with
  non-{{Abelian}} groups},}\ }\href {https://doi.org/10.1063/5.0219434}
  {\bibfield  {journal} {\bibinfo  {journal} {J. Chem. Phys.}\ }\textbf
  {\bibinfo {volume} {161}},\ \bibinfo {pages} {044102} (\bibinfo {year}
  {2024})}\BibitemShut {NoStop}%
\bibitem [{\citenamefont {Zhang}, \citenamefont {Wang},\ and\ \citenamefont
  {Wang}(2024)}]{Neural2024zhang}%
  \BibitemOpen
  \bibfield  {author} {\bibinfo {author} {\bibfnamefont {Q.}~\bibnamefont
  {Zhang}}, \bibinfo {author} {\bibfnamefont {R.-S.}\ \bibnamefont {Wang}},\
  and\ \bibinfo {author} {\bibfnamefont {L.}~\bibnamefont {Wang}},\ }\bibfield
  {title} {\enquote {\bibinfo {title} {Neural canonical transformations for
  vibrational spectra of molecules},}\ }\href
  {https://doi.org/10.1063/5.0209255} {\bibfield  {journal} {\bibinfo
  {journal} {J. Chem. Phys.}\ }\textbf {\bibinfo {volume} {161}},\ \bibinfo
  {pages} {024103} (\bibinfo {year} {2024})}\BibitemShut {NoStop}%
\bibitem [{\citenamefont {Solomon}\ \emph {et~al.}(1971)\citenamefont
  {Solomon}, \citenamefont {Jefferts}, \citenamefont {Penzias},\ and\
  \citenamefont {Wilson}}]{Detection1971solomon}%
  \BibitemOpen
  \bibfield  {author} {\bibinfo {author} {\bibfnamefont {P.~M.}\ \bibnamefont
  {Solomon}}, \bibinfo {author} {\bibfnamefont {K.~B.}\ \bibnamefont
  {Jefferts}}, \bibinfo {author} {\bibfnamefont {A.~A.}\ \bibnamefont
  {Penzias}},\ and\ \bibinfo {author} {\bibfnamefont {R.~W.}\ \bibnamefont
  {Wilson}},\ }\bibfield  {title} {\enquote {\bibinfo {title} {Detection of
  {{Millimeter Emission Lines}} from {{Interstellar Methyl Cyanide}}},}\ }\href
  {https://doi.org/10.1086/180794} {\bibfield  {journal} {\bibinfo  {journal}
  {ApJ}\ }\textbf {\bibinfo {volume} {168}},\ \bibinfo {pages} {L107} (\bibinfo
  {year} {1971})}\BibitemShut {NoStop}%
\bibitem [{\citenamefont {Lobert}\ \emph {et~al.}(1990)\citenamefont {Lobert},
  \citenamefont {Scharffe}, \citenamefont {Hao},\ and\ \citenamefont
  {Crutzen}}]{Importance1990lobert}%
  \BibitemOpen
  \bibfield  {author} {\bibinfo {author} {\bibfnamefont {J.~M.}\ \bibnamefont
  {Lobert}}, \bibinfo {author} {\bibfnamefont {D.~H.}\ \bibnamefont
  {Scharffe}}, \bibinfo {author} {\bibfnamefont {W.~M.}\ \bibnamefont {Hao}},\
  and\ \bibinfo {author} {\bibfnamefont {P.~J.}\ \bibnamefont {Crutzen}},\
  }\bibfield  {title} {\enquote {\bibinfo {title} {Importance of biomass
  burning in the atmospheric budgets of nitrogen-containing gases},}\ }\href
  {https://doi.org/10.1038/346552a0} {\bibfield  {journal} {\bibinfo  {journal}
  {Nature}\ }\textbf {\bibinfo {volume} {346}},\ \bibinfo {pages} {552--554}
  (\bibinfo {year} {1990})}\BibitemShut {NoStop}%
\bibitem [{\citenamefont {O'Leary}\ \emph {et~al.}(2012)\citenamefont
  {O'Leary}, \citenamefont {Ruth}, \citenamefont {Dixneuf}, \citenamefont
  {Orphal},\ and\ \citenamefont {Varma}}]{Infrared2012oleary}%
  \BibitemOpen
  \bibfield  {author} {\bibinfo {author} {\bibfnamefont {D.~M.}\ \bibnamefont
  {O'Leary}}, \bibinfo {author} {\bibfnamefont {A.~A.}\ \bibnamefont {Ruth}},
  \bibinfo {author} {\bibfnamefont {S.}~\bibnamefont {Dixneuf}}, \bibinfo
  {author} {\bibfnamefont {J.}~\bibnamefont {Orphal}},\ and\ \bibinfo {author}
  {\bibfnamefont {R.}~\bibnamefont {Varma}},\ }\bibfield  {title} {\enquote
  {\bibinfo {title} {The near infrared cavity-enhanced absorption spectrum of
  methyl cyanide},}\ }\href {https://doi.org/10.1016/j.jqsrt.2012.02.022}
  {\bibfield  {journal} {\bibinfo  {journal} {J. Quant. Spectrosc. Radiat.
  Transf.}\ }\textbf {\bibinfo {volume} {113}},\ \bibinfo {pages} {1138--1147}
  (\bibinfo {year} {2012})}\BibitemShut {NoStop}%
\bibitem [{\citenamefont {Zhao}\ and\ \citenamefont
  {Wright}(1999)}]{Measurement1999zhao}%
  \BibitemOpen
  \bibfield  {author} {\bibinfo {author} {\bibfnamefont {W.}~\bibnamefont
  {Zhao}}\ and\ \bibinfo {author} {\bibfnamefont {J.~C.}\ \bibnamefont
  {Wright}},\ }\bibfield  {title} {\enquote {\bibinfo {title} {Measurement of
  {$\chi$}{\textsuperscript{(3)}} for {{Doubly Vibrationally Enhanced Four Wave
  Mixing Spectroscopy}}},}\ }\href
  {https://doi.org/10.1103/PhysRevLett.83.1950} {\bibfield  {journal} {\bibinfo
   {journal} {Phys. Rev. Lett.}\ }\textbf {\bibinfo {volume} {83}},\ \bibinfo
  {pages} {1950--1953} (\bibinfo {year} {1999})}\BibitemShut {NoStop}%
\bibitem [{\citenamefont {McDonnell}\ \emph {et~al.}(2024)\citenamefont
  {McDonnell}, \citenamefont {Oram}, \citenamefont {Boyer}, \citenamefont
  {Kohler}, \citenamefont {Meyer}, \citenamefont {Sibert~Iii},\ and\
  \citenamefont {Wright}}]{Direct2024mcdonnell}%
  \BibitemOpen
  \bibfield  {author} {\bibinfo {author} {\bibfnamefont {R.~P.}\ \bibnamefont
  {McDonnell}}, \bibinfo {author} {\bibfnamefont {K.}~\bibnamefont {Oram}},
  \bibinfo {author} {\bibfnamefont {M.~A.}\ \bibnamefont {Boyer}}, \bibinfo
  {author} {\bibfnamefont {D.~D.}\ \bibnamefont {Kohler}}, \bibinfo {author}
  {\bibfnamefont {K.~A.}\ \bibnamefont {Meyer}}, \bibinfo {author}
  {\bibfnamefont {E.~L.}\ \bibnamefont {Sibert~Iii}},\ and\ \bibinfo {author}
  {\bibfnamefont {J.~C.}\ \bibnamefont {Wright}},\ }\bibfield  {title}
  {\enquote {\bibinfo {title} {Direct {{Probe}} of {{Vibrational Fingerprint}}
  and {{Combination Band Coupling}}},}\ }\href
  {https://doi.org/10.1021/acs.jpclett.4c00297} {\bibfield  {journal} {\bibinfo
   {journal} {J. Phys. Chem. Lett.}\ }\textbf {\bibinfo {volume} {15}},\
  \bibinfo {pages} {3975--3981} (\bibinfo {year} {2024})}\BibitemShut {NoStop}%
\bibitem [{\citenamefont {Westermann}\ and\ \citenamefont
  {Manthe}(2012)}]{First2012westermann}%
  \BibitemOpen
  \bibfield  {author} {\bibinfo {author} {\bibfnamefont {T.}~\bibnamefont
  {Westermann}}\ and\ \bibinfo {author} {\bibfnamefont {U.}~\bibnamefont
  {Manthe}},\ }\bibfield  {title} {\enquote {\bibinfo {title} {First principle
  nonlinear quantum dynamics using a correlation-based von {{Neumann}}
  entropy},}\ }\href {https://doi.org/10.1063/1.4720567} {\bibfield  {journal}
  {\bibinfo  {journal} {J. Chem. Phys.}\ }\textbf {\bibinfo {volume} {136}},\
  \bibinfo {pages} {204116} (\bibinfo {year} {2012})}\BibitemShut {NoStop}%
\bibitem [{\citenamefont {Glaser}\ \emph {et~al.}(2024)\citenamefont {Glaser},
  \citenamefont {Baiardi}, \citenamefont {Lieberherr},\ and\ \citenamefont
  {Reiher}}]{Vibrational2024glaser}%
  \BibitemOpen
  \bibfield  {author} {\bibinfo {author} {\bibfnamefont {N.}~\bibnamefont
  {Glaser}}, \bibinfo {author} {\bibfnamefont {A.}~\bibnamefont {Baiardi}},
  \bibinfo {author} {\bibfnamefont {A.~Z.}\ \bibnamefont {Lieberherr}},\ and\
  \bibinfo {author} {\bibfnamefont {M.}~\bibnamefont {Reiher}},\ }\bibfield
  {title} {\enquote {\bibinfo {title} {Vibrational {{Entanglement}} through the
  {{Lens}} of {{Quantum Information Measures}}},}\ }\href
  {https://doi.org/10.1021/acs.jpclett.4c01298} {\bibfield  {journal} {\bibinfo
   {journal} {J. Phys. Chem. Lett.}\ ,\ \bibinfo {pages} {6958--6965}}
  (\bibinfo {year} {2024})}\BibitemShut {NoStop}%
\bibitem [{\citenamefont {White}(1992)}]{Density1992white}%
  \BibitemOpen
  \bibfield  {author} {\bibinfo {author} {\bibfnamefont {S.~R.}\ \bibnamefont
  {White}},\ }\bibfield  {title} {\enquote {\bibinfo {title} {Density matrix
  formulation for quantum renormalization groups},}\ }\href
  {https://doi.org/10.1103/PhysRevLett.69.2863} {\bibfield  {journal} {\bibinfo
   {journal} {Phys. Rev. Lett.}\ }\textbf {\bibinfo {volume} {69}},\ \bibinfo
  {pages} {2863--2866} (\bibinfo {year} {1992})}\BibitemShut {NoStop}%
\bibitem [{\citenamefont {Or{\'u}s}(2014)}]{Practical2014orus}%
  \BibitemOpen
  \bibfield  {author} {\bibinfo {author} {\bibfnamefont {R.}~\bibnamefont
  {Or{\'u}s}},\ }\bibfield  {title} {\enquote {\bibinfo {title} {A practical
  introduction to tensor networks: {{Matrix}} product states and projected
  entangled pair states},}\ }\href {https://doi.org/10.1016/j.aop.2014.06.013}
  {\bibfield  {journal} {\bibinfo  {journal} {Ann. Phys.}\ }\textbf {\bibinfo
  {volume} {349}},\ \bibinfo {pages} {117--158} (\bibinfo {year}
  {2014})}\BibitemShut {NoStop}%
\bibitem [{\citenamefont {Meyer}(2012)}]{Studying2012meyer}%
  \BibitemOpen
  \bibfield  {author} {\bibinfo {author} {\bibfnamefont {H.-D.}\ \bibnamefont
  {Meyer}},\ }\bibfield  {title} {\enquote {\bibinfo {title} {Studying
  molecular quantum dynamics with the multiconfiguration time-dependent
  {{Hartree}} method},}\ }\href {https://doi.org/10.1002/wcms.87} {\bibfield
  {journal} {\bibinfo  {journal} {Wiley Interdiscip. Rev. Comput. Mol. Sci.}\
  }\textbf {\bibinfo {volume} {2}},\ \bibinfo {pages} {351--374} (\bibinfo
  {year} {2012})}\BibitemShut {NoStop}%
\bibitem [{\citenamefont {Wang}(2015)}]{Multilayer2015wang}%
  \BibitemOpen
  \bibfield  {author} {\bibinfo {author} {\bibfnamefont {H.}~\bibnamefont
  {Wang}},\ }\bibfield  {title} {\enquote {\bibinfo {title} {Multilayer
  {{Multiconfiguration Time-Dependent Hartree Theory}}},}\ }\href
  {https://doi.org/10.1021/acs.jpca.5b03256} {\bibfield  {journal} {\bibinfo
  {journal} {J. Phys. Chem. A}\ }\textbf {\bibinfo {volume} {119}},\ \bibinfo
  {pages} {7951--7965} (\bibinfo {year} {2015})}\BibitemShut {NoStop}%
\bibitem [{\citenamefont {Manthe}(2017)}]{Wavepacket2017manthe}%
  \BibitemOpen
  \bibfield  {author} {\bibinfo {author} {\bibfnamefont {U.}~\bibnamefont
  {Manthe}},\ }\bibfield  {title} {\enquote {\bibinfo {title} {Wavepacket
  dynamics and the multi-configurational time-dependent {{Hartree}}
  approach},}\ }\href {https://doi.org/10.1088/1361-648X/aa6e96} {\bibfield
  {journal} {\bibinfo  {journal} {J. Phys. Condens. Matter}\ }\textbf {\bibinfo
  {volume} {29}},\ \bibinfo {pages} {253001} (\bibinfo {year}
  {2017})}\BibitemShut {NoStop}%
\bibitem [{\citenamefont {Shavitt}\ \emph {et~al.}(1973)\citenamefont
  {Shavitt}, \citenamefont {Bender}, \citenamefont {Pipano},\ and\
  \citenamefont {Hosteny}}]{Iterative1973shavitt}%
  \BibitemOpen
  \bibfield  {author} {\bibinfo {author} {\bibfnamefont {I.}~\bibnamefont
  {Shavitt}}, \bibinfo {author} {\bibfnamefont {C.~F.}\ \bibnamefont {Bender}},
  \bibinfo {author} {\bibfnamefont {A.}~\bibnamefont {Pipano}},\ and\ \bibinfo
  {author} {\bibfnamefont {R.~P.}\ \bibnamefont {Hosteny}},\ }\bibfield
  {title} {\enquote {\bibinfo {title} {The iterative calculation of several of
  the lowest or highest eigenvalues and corresponding eigenvectors of very
  large symmetric matrices},}\ }\href
  {https://doi.org/10.1016/0021-9991(73)90149-6} {\bibfield  {journal}
  {\bibinfo  {journal} {J. Comput. Phys.}\ }\textbf {\bibinfo {volume} {11}},\
  \bibinfo {pages} {90--108} (\bibinfo {year} {1973})}\BibitemShut {NoStop}%
\bibitem [{\citenamefont {Larsson}\ and\ \citenamefont
  {Viel}(2024)}]{vibronic2024larsson}%
  \BibitemOpen
  \bibfield  {author} {\bibinfo {author} {\bibfnamefont {H.~R.}\ \bibnamefont
  {Larsson}}\ and\ \bibinfo {author} {\bibfnamefont {A.}~\bibnamefont {Viel}},\
  }\bibfield  {title} {\enquote {\bibinfo {title} {2500 vibronic eigenstates of
  the {{NO}}{\textsubscript{3}} radical},}\ }\href
  {https://doi.org/10.1039/D4CP02653E} {\bibfield  {journal} {\bibinfo
  {journal} {Phys. Chem. Chem. Phys.}\ }\textbf {\bibinfo {volume} {26}},\
  \bibinfo {pages} {24506--24523} (\bibinfo {year} {2024})}\BibitemShut
  {NoStop}%
\bibitem [{\citenamefont {Zhai}\ \emph {et~al.}(2023)\citenamefont {Zhai},
  \citenamefont {Larsson}, \citenamefont {Lee}, \citenamefont {Cui},
  \citenamefont {Zhu}, \citenamefont {Sun}, \citenamefont {Peng}, \citenamefont
  {Peng}, \citenamefont {Liao}, \citenamefont {T{\"o}lle}, \citenamefont
  {Yang}, \citenamefont {Li},\ and\ \citenamefont {Chan}}]{Block22023zhai}%
  \BibitemOpen
  \bibfield  {author} {\bibinfo {author} {\bibfnamefont {H.}~\bibnamefont
  {Zhai}}, \bibinfo {author} {\bibfnamefont {H.~R.}\ \bibnamefont {Larsson}},
  \bibinfo {author} {\bibfnamefont {S.}~\bibnamefont {Lee}}, \bibinfo {author}
  {\bibfnamefont {Z.-H.}\ \bibnamefont {Cui}}, \bibinfo {author} {\bibfnamefont
  {T.}~\bibnamefont {Zhu}}, \bibinfo {author} {\bibfnamefont {C.}~\bibnamefont
  {Sun}}, \bibinfo {author} {\bibfnamefont {L.}~\bibnamefont {Peng}}, \bibinfo
  {author} {\bibfnamefont {R.}~\bibnamefont {Peng}}, \bibinfo {author}
  {\bibfnamefont {K.}~\bibnamefont {Liao}}, \bibinfo {author} {\bibfnamefont
  {J.}~\bibnamefont {T{\"o}lle}}, \bibinfo {author} {\bibfnamefont
  {J.}~\bibnamefont {Yang}}, \bibinfo {author} {\bibfnamefont {S.}~\bibnamefont
  {Li}},\ and\ \bibinfo {author} {\bibfnamefont {G.~K.-L.}\ \bibnamefont
  {Chan}},\ }\bibfield  {title} {\enquote {\bibinfo {title} {Block2: {{A}}
  comprehensive open source framework to develop and apply state-of-the-art
  {{DMRG}} algorithms in electronic structure and beyond},}\ }\href
  {https://doi.org/10.1063/5.0180424} {\bibfield  {journal} {\bibinfo
  {journal} {J. Chem. Phys.}\ }\textbf {\bibinfo {volume} {159}},\ \bibinfo
  {pages} {234801} (\bibinfo {year} {2023})}\BibitemShut {NoStop}%
\bibitem [{\citenamefont {Light}\ and\ \citenamefont
  {Carrington~Jr}(2000)}]{Discretevariable2000light}%
  \BibitemOpen
  \bibfield  {author} {\bibinfo {author} {\bibfnamefont {J.~C.}\ \bibnamefont
  {Light}}\ and\ \bibinfo {author} {\bibfnamefont {T.}~\bibnamefont
  {Carrington~Jr}},\ }\bibfield  {title} {\enquote {\bibinfo {title}
  {Discrete-variable representations and their utilization},}\ }\href
  {https://doi.org/10.1002/9780470141731.ch4} {\bibfield  {journal} {\bibinfo
  {journal} {Adv. Chem. Phys.}\ }\textbf {\bibinfo {volume} {114}},\ \bibinfo
  {pages} {263--310} (\bibinfo {year} {2000})}\BibitemShut {NoStop}%
\bibitem [{\citenamefont {Tannor}\ \emph {et~al.}(2018)\citenamefont {Tannor},
  \citenamefont {Machnes}, \citenamefont {Ass{\'e}mat},\ and\ \citenamefont
  {Larsson}}]{PhaseSpace2018tannor}%
  \BibitemOpen
  \bibfield  {author} {\bibinfo {author} {\bibfnamefont {D.}~\bibnamefont
  {Tannor}}, \bibinfo {author} {\bibfnamefont {S.}~\bibnamefont {Machnes}},
  \bibinfo {author} {\bibfnamefont {E.}~\bibnamefont {Ass{\'e}mat}},\ and\
  \bibinfo {author} {\bibfnamefont {H.~R.}\ \bibnamefont {Larsson}},\
  }\bibfield  {title} {\enquote {\bibinfo {title} {Phase-{{Space Versus
  Coordinate-Space Methods}}: {{Prognosis}} for {{Large Quantum
  Calculations}}},}\ }in\ \href {https://doi.org/10.1002/9781119374978.ch10}
  {\emph {\bibinfo {booktitle} {Advances in {{Chemical Physics}}}}},\ Vol.\
  \bibinfo {volume} {163},\ \bibinfo {editor} {edited by\ \bibinfo {editor}
  {\bibfnamefont {K.~B.}\ \bibnamefont {Whaley}}}\ (\bibinfo  {publisher} {John
  Wiley \& Sons, Inc.},\ \bibinfo {address} {Hoboken, NJ, USA},\ \bibinfo
  {year} {2018})\ pp.\ \bibinfo {pages} {273--323}\BibitemShut {NoStop}%
\bibitem [{\citenamefont {Larsson}, \citenamefont {Hartke},\ and\ \citenamefont
  {Tannor}(2016)}]{Efficient2016larsson}%
  \BibitemOpen
  \bibfield  {author} {\bibinfo {author} {\bibfnamefont {H.~R.}\ \bibnamefont
  {Larsson}}, \bibinfo {author} {\bibfnamefont {B.}~\bibnamefont {Hartke}},\
  and\ \bibinfo {author} {\bibfnamefont {D.~J.}\ \bibnamefont {Tannor}},\
  }\bibfield  {title} {\enquote {\bibinfo {title} {Efficient molecular quantum
  dynamics in coordinate and phase space using pruned bases},}\ }\href
  {https://doi.org/http://dx.doi.org/10.1063/1.4967432} {\bibfield  {journal}
  {\bibinfo  {journal} {J. Chem. Phys.}\ }\textbf {\bibinfo {volume} {145}},\
  \bibinfo {pages} {204108} (\bibinfo {year} {2016})}\BibitemShut {NoStop}%
\bibitem [{\citenamefont {Larsson}\ and\ \citenamefont
  {Tannor}(2017)}]{Dynamical2017larsson}%
  \BibitemOpen
  \bibfield  {author} {\bibinfo {author} {\bibfnamefont {H.~R.}\ \bibnamefont
  {Larsson}}\ and\ \bibinfo {author} {\bibfnamefont {D.~J.}\ \bibnamefont
  {Tannor}},\ }\bibfield  {title} {\enquote {\bibinfo {title} {Dynamical
  pruning of the multiconfiguration time-dependent {{Hartree}} ({{DP-MCTDH}})
  method: {{An}} efficient approach for multidimensional quantum dynamics},}\
  }\href {https://doi.org/10.1063/1.4993219} {\bibfield  {journal} {\bibinfo
  {journal} {J. Chem. Phys.}\ }\textbf {\bibinfo {volume} {147}},\ \bibinfo
  {pages} {044103} (\bibinfo {year} {2017})}\BibitemShut {NoStop}%
\bibitem [{\citenamefont {Tagliacozzo}, \citenamefont {Evenbly},\ and\
  \citenamefont {Vidal}(2009)}]{Simulation2009tagliacozzo}%
  \BibitemOpen
  \bibfield  {author} {\bibinfo {author} {\bibfnamefont {L.}~\bibnamefont
  {Tagliacozzo}}, \bibinfo {author} {\bibfnamefont {G.}~\bibnamefont
  {Evenbly}},\ and\ \bibinfo {author} {\bibfnamefont {G.}~\bibnamefont
  {Vidal}},\ }\bibfield  {title} {\enquote {\bibinfo {title} {Simulation of
  two-dimensional quantum systems using a tree tensor network that exploits the
  entropic area law},}\ }\href {https://doi.org/10.1103/PhysRevB.80.235127}
  {\bibfield  {journal} {\bibinfo  {journal} {Phys. Rev. B}\ }\textbf {\bibinfo
  {volume} {80}},\ \bibinfo {pages} {235127} (\bibinfo {year}
  {2009})}\BibitemShut {NoStop}%
\bibitem [{\citenamefont {Hubig}, \citenamefont {Haegeman},\ and\ \citenamefont
  {Schollw{\"o}ck}(2018)}]{Error2018hubig}%
  \BibitemOpen
  \bibfield  {author} {\bibinfo {author} {\bibfnamefont {C.}~\bibnamefont
  {Hubig}}, \bibinfo {author} {\bibfnamefont {J.}~\bibnamefont {Haegeman}},\
  and\ \bibinfo {author} {\bibfnamefont {U.}~\bibnamefont {Schollw{\"o}ck}},\
  }\bibfield  {title} {\enquote {\bibinfo {title} {Error estimates for
  extrapolations with matrix-product states},}\ }\href
  {https://doi.org/10.1103/PhysRevB.97.045125} {\bibfield  {journal} {\bibinfo
  {journal} {Phys. Rev. B}\ }\textbf {\bibinfo {volume} {97}},\ \bibinfo
  {pages} {045125} (\bibinfo {year} {2018})}\BibitemShut {NoStop}%
\bibitem [{\citenamefont {Henry}\ and\ \citenamefont
  {Amat}(1961)}]{Cubic1961henry}%
  \BibitemOpen
  \bibfield  {author} {\bibinfo {author} {\bibfnamefont {L.}~\bibnamefont
  {Henry}}\ and\ \bibinfo {author} {\bibfnamefont {G.}~\bibnamefont {Amat}},\
  }\bibfield  {title} {\enquote {\bibinfo {title} {The cubic anharmonic
  potential function of polyatomic molecules},}\ }\href
  {https://doi.org/10.1016/0022-2852(61)90096-0} {\bibfield  {journal}
  {\bibinfo  {journal} {J. Mol. Spectrosc.}\ }\textbf {\bibinfo {volume} {5}},\
  \bibinfo {pages} {319--325} (\bibinfo {year} {1961})}\BibitemShut {NoStop}%
\bibitem [{\citenamefont {Henry}\ and\ \citenamefont
  {Amat}(1965)}]{Quartic1965henry}%
  \BibitemOpen
  \bibfield  {author} {\bibinfo {author} {\bibfnamefont {L.}~\bibnamefont
  {Henry}}\ and\ \bibinfo {author} {\bibfnamefont {G.}~\bibnamefont {Amat}},\
  }\bibfield  {title} {\enquote {\bibinfo {title} {The quartic anharmonic
  potential function of polyatomic molecules},}\ }\href
  {https://doi.org/10.1016/0022-2852(65)90034-2} {\bibfield  {journal}
  {\bibinfo  {journal} {J. Mol. Spectrosc.}\ }\textbf {\bibinfo {volume}
  {15}},\ \bibinfo {pages} {168--179} (\bibinfo {year} {1965})}\BibitemShut
  {NoStop}%
\bibitem [{\citenamefont {Larsson}\ \emph
  {et~al.}(2022{\natexlab{c}})\citenamefont {Larsson}, \citenamefont {Zhai},
  \citenamefont {Gunst},\ and\ \citenamefont {Chan}}]{Matrix2022larsson}%
  \BibitemOpen
  \bibfield  {author} {\bibinfo {author} {\bibfnamefont {H.~R.}\ \bibnamefont
  {Larsson}}, \bibinfo {author} {\bibfnamefont {H.}~\bibnamefont {Zhai}},
  \bibinfo {author} {\bibfnamefont {K.}~\bibnamefont {Gunst}},\ and\ \bibinfo
  {author} {\bibfnamefont {G.~K.-L.}\ \bibnamefont {Chan}},\ }\bibfield
  {title} {\enquote {\bibinfo {title} {Matrix {{Product States}} with {{Large
  Sites}}},}\ }\href {https://doi.org/10.1021/acs.jctc.1c00957} {\bibfield
  {journal} {\bibinfo  {journal} {J. Chem. Theory Comput.}\ }\textbf {\bibinfo
  {volume} {18}},\ \bibinfo {pages} {749--762} (\bibinfo {year}
  {2022}{\natexlab{c}})}\BibitemShut {NoStop}%
\bibitem [{\citenamefont {Ndong}\ \emph {et~al.}(2012)\citenamefont {Ndong},
  \citenamefont {{Joubert-Doriol}}, \citenamefont {Meyer}, \citenamefont
  {Nauts}, \citenamefont {Gatti},\ and\ \citenamefont
  {Lauvergnat}}]{Automatic2012ndong}%
  \BibitemOpen
  \bibfield  {author} {\bibinfo {author} {\bibfnamefont {M.}~\bibnamefont
  {Ndong}}, \bibinfo {author} {\bibfnamefont {L.}~\bibnamefont
  {{Joubert-Doriol}}}, \bibinfo {author} {\bibfnamefont {H.-D.}\ \bibnamefont
  {Meyer}}, \bibinfo {author} {\bibfnamefont {A.}~\bibnamefont {Nauts}},
  \bibinfo {author} {\bibfnamefont {F.}~\bibnamefont {Gatti}},\ and\ \bibinfo
  {author} {\bibfnamefont {D.}~\bibnamefont {Lauvergnat}},\ }\bibfield  {title}
  {\enquote {\bibinfo {title} {Automatic computer procedure for generating
  exact and analytical kinetic energy operators based on the polyspherical
  approach},}\ }\href {https://doi.org/10.1063/1.3675163} {\bibfield  {journal}
  {\bibinfo  {journal} {J. Chem. Phys.}\ }\textbf {\bibinfo {volume} {136}},\
  \bibinfo {pages} {034107} (\bibinfo {year} {2012})}\BibitemShut {NoStop}%
\bibitem [{\citenamefont {Schr{\"o}der}(2020)}]{Transforming2020schroder}%
  \BibitemOpen
  \bibfield  {author} {\bibinfo {author} {\bibfnamefont {M.}~\bibnamefont
  {Schr{\"o}der}},\ }\bibfield  {title} {\enquote {\bibinfo {title}
  {Transforming high-dimensional potential energy surfaces into a canonical
  polyadic decomposition using {{Monte Carlo}} methods},}\ }\href
  {https://doi.org/10.1063/1.5140085} {\bibfield  {journal} {\bibinfo
  {journal} {J. Chem. Phys.}\ }\textbf {\bibinfo {volume} {152}},\ \bibinfo
  {pages} {024108} (\bibinfo {year} {2020})}\BibitemShut {NoStop}%
\end{thebibliography}
\end{document}